\documentclass[a4paper,12pt]{article}
\usepackage{jheppub_modified_green}

\usepackage{placeins}

\usepackage{epsfig}
\usepackage[makeroom]{cancel}
\usepackage{amsfonts}
\usepackage{amssymb,esint}
\usepackage{enumitem}
\usepackage{comment}
\usepackage[normalem]{ulem}
\usepackage{amsthm}
\usepackage{subfig}
\usepackage{bm}
\usepackage{hyperref}
\usepackage{mathrsfs}
\usepackage{bbm}
\usepackage{cancel}
\usepackage{xcolor}
\usepackage{relsize}
\usepackage{float, slashed, graphicx, amssymb, amsmath}
\usepackage{shuffle}
\usepackage{pgfplots}
\usepackage{tikz-feynman}
\pgfplotsset{compat=1.18}
\tikzfeynmanset{compat=1.1.0}
\tikzfeynmanset{graviton/.style={circle, draw=green!60, fill=green!5, very thick, minimum size=7mm}}
\tikzfeynmanset{codot/.style={/tikz/shape=circle,/tikz/fill=white,/tikz/minimum size=0.1cm,/tikz/inner sep=1.8pt}}
\tikzfeynmanset{myblob/.style={/tikz/shape=rectangle,/tikz/fill=red,/tikz/minimum size=0.2cm,/tikz/inner sep=1.8pt} }
\tikzfeynmanset{ghc/.style={/tikz/shape=circle,/tikz/fill=white,/tikz/minimum size=0.05cm,} }
\tikzfeynmanset{HV/.style=
{/tikz/shape=circle,
/tikz/fill={rgb:black,1;white,2},
 /tikz/minimum size=0.01cm,/tikz/inner sep=3.0pt
 } }
\tikzfeynmanset{GR/.style={/tikz/shape=ellipse,/tikz/fill={rgb:black,1;white,2},/tikz/minimum size=0.3cm,} }
\tikzfeynmanset{GR2/.style={/tikz/shape=ellipse,/tikz/fill={rgb:black,1;white,2},/tikz/minimum width = 1.2cm, 
    /tikz/minimum height = 3.4cm} }
\tikzset{box/.pic={\filldraw[fill=black]  (0,0) circle (2.5pt); \filldraw [fill=black] (0.5,0) circle (2.5pt); \draw [line width=5pt] (0,0) -- (0.5,0);}}
 \tikzfeynmanset{myblob2/.style=
{/tikz/shape=rectangle,
/tikz/fill=black,
 /tikz/minimum width=0.1cm,/tikz/inner sep=1.8pt
 } }
 \tikzfeynmanset{sb/.style=
{/tikz/shape=circle,
/tikz/fill=black,
 /tikz/minimum size=0.3cm,/tikz/inner sep=1.pt
 } }
 \tikzfeynmanset{myblob/.style=
{/tikz/shape=ellipse,
/tikz/fill=red,
 /tikz/minimum width=0.5cm,
 } }

\usetikzlibrary{arrows.meta} % decent arrow in the graphs
\usetikzlibrary{calc}
\usetikzlibrary{decorations.pathmorphing}
\usetikzlibrary{decorations.pathreplacing}
\usetikzlibrary{decorations.markings}
\tikzset{
   vector2/.style={decorate, decoration={snake, amplitude=1pt, segment length=6pt}, draw,double},
   vector/.style={decorate, decoration={snake, amplitude=1pt, segment length=6pt}, draw},
	provector/.style={decorate, decoration={snake,amplitude=2.5pt}, draw},
	antivector/.style={decorate, decoration={snake,amplitude=-2.5pt}, draw},
    fermion/.style={draw=black, postaction={decorate},
        decoration={markings,mark=at position .55 with {\arrow[draw=black]{>}}}},
    fermionbar/.style={draw=black, postaction={decorate},
        decoration={markings,mark=at position .55 with {\arrow[draw=black]{<}}}},
    fermionnoarrow/.style={draw=black},
    gluon/.style={decorate, draw=black,
        decoration={coil,amplitude=4pt, segment length=5pt}},
    scalar/.style={dashed,draw=black, postaction={decorate},
        decoration={markings,mark=at position .55 with {\arrow[draw=black]{>}}}},
    scalarbar/.style={dashed,draw=black, postaction={decorate},
        decoration={markings,mark=at position .55 with {\arrow[draw=black]{<}}}},
    scalarnoarrow/.style={dashed,draw=black},
    electron/.style={draw=black, postaction={decorate},
        decoration={markings,mark=at position .55 with {\arrow[draw=black]{>}}}},
	bigvector/.style={decorate, decoration={snake,amplitude=4pt}, draw},
}
\tikzset{cross/.style={cross out, draw, 
         minimum size=2*(#1-\pgflinewidth), 
         inner sep=0pt, outer sep=0pt}}

\tikzstyle{block} = [draw, rectangle, 
    minimum height=3em, minimum width=6em]

\usepackage[T1]{fontenc} % if needed

\def\sc#1{\overline{#1}}
\makeatletter
\newcommand \UPlus {\mathop {\operator@font \uplus }\limits }
\makeatother
\makeatletter
\newcommand \Bigcup {\mathop {\operator@font \bigcup }\limits }
\makeatother
\def\LabelNote#1{}%\smash{\hbox to\phipt{\raise1ex\hbox{\tiny[#1]}\hss}}}
\def\Label#1{\label{#1}%
\smash{\hbox to\phipt{\raise1ex\hbox{\tiny[#1]}\hss}}}

\def\mdot{{\cdot}}

\definecolor{bananayellow}{rgb}{1.0, 0.88, 0.21}
\definecolor{amber}{rgb}{1.0, 0.75, 0.0}

\newcommand{\cI}{\mathcal{I}}
\newcommand{\cM}{\mathcal{M}}

\newcommand{\cC}{\mathcal{C}}

\newcommand{\bt}{\tilde{b}}
\newcommand{\at}{\tilde{a}}
\newcommand{\fb}{\mathfrak{b}}
\newcommand{\fu}{\mathfrak{u}}

\newcommand{\cD}{\mathcal{D}}

\newcommand{\eps}{\epsilon}
\newcommand{\vareps}{\varepsilon}

\newcommand{\pb}{\bar p}

\def\nn{\nonumber}

\def\spa#1.#2{\left\langle#1\,#2\right\rangle}
\def\spb#1.#2{\left[#1\,#2\right]}

\def\be{\begin{equation}}
\def\ee{\end{equation}}
\def\bea{\begin{eqnarray}}
\def\eea{\end{eqnarray}}  
\allowdisplaybreaks

\usepackage{amssymb,amsmath}
\usepackage{mathtools} %added for coloneqq
\usepackage{cancel} %crossing-out terms
\usepackage{graphicx} %Lets us rotate symbols

\usepackage{hyperref}
\hypersetup{
	colorlinks=true,
	linktoc=page,
	citecolor=americanrose,
	linkcolor=cadmiumgreen,
	urlcolor=blue} 
\definecolor{americanrose}{rgb}{1.0, 0.01, 0.24}
\definecolor{cadmiumgreen}{rgb}{0.0, 0.42, 0.24}

\usepackage[colorinlistoftodos]{todonotes}
\newcommand{\Cdot}{{\cdot}} 
\makeatletter
\newcommand*{\bigcdot}{}% Check if undefined
\DeclareRobustCommand*{\bigcdot}{%
  \mathbin{\mathpalette\bigcdot@{}}%
}
\newcommand*{\bigcdot@scalefactor}{.6}
\newcommand*{\bigcdot@widthfactor}{1.25}
\newcommand*{\bigcdot@}[2]{%
  \sbox0{$#1\vcenter{}$}% math axis
  \sbox2{$#1\cdot\m@th$}%
  \hbox to \bigcdot@widthfactor\wd2{%
    \hfil
    \raise\ht0\hbox{%
      \scalebox{\bigcdot@scalefactor}{%
        \lower\ht0\hbox{$#1\bullet\m@th$}%
      }%
    }%
    \hfil
  }%
}
\makeatother
\def\nn{\nonumber}

\title{Spinning  waveforms in cubic effective field theories of gravity 
}
\author{Andreas Brandhuber$\mbox{}^{a}$,}
\author{Graham R.~Brown$\mbox{}^{a}$,}
\author{Gang Chen$\mbox{}^{b}$,\\}
\author{\hspace{-0.2cm}Gabriele Travaglini$\mbox{}^{a}$}
\author{and Pablo Vives Matasan$\mbox{}^{a}$}
\affiliation{$\mbox{}^{a}$Centre for Theoretical Physics, Department of Physics and Astronomy, \\
Queen Mary University of London, Mile End Road, London E1 4NS, United Kingdom}
\affiliation{$\mbox{}^{b}$Niels Bohr International Academy,
Niels Bohr Institute, University of Copenhagen,\\
Blegdamsvej 17, DK-2100 Copenhagen \O, Denmark}
\emailAdd{a.brandhuber@qmul.ac.uk}
\emailAdd{graham.brown@qmul.ac.uk}
\emailAdd{gang.chen@nbi.ku.dk}
\emailAdd{g.travaglini@qmul.ac.uk}
\emailAdd{p.vivesmatasan@qmul.ac.uk}
\begin{document}
\begin{flushright}
	QMUL-PH-24-11
\end{flushright}

\abstract{
We derive analytic all-order-in-spin expressions  for  the leading-order time-domain waveforms  generated in the scattering of two Kerr black holes with arbitrary masses and spin vectors in the presence of all independent cubic deformations of Einstein-Hilbert gravity. These are  the  two parity-even interactions $I_1$ and $G_3$, and  the parity-odd ones 
$\tilde{I}_1$ and $\tilde{G}_3$. 
Our results are obtained using three independent methods: a particularly efficient  direct integration and tensor reduction approach; integration by parts combined with  the method of differential equations; and finally a residue computation. 
For the case of the $G_3$ and $\tilde{G}_3$ deformations we can express the spinning waveform  in terms of the scalar waveform  with appropriately shifted impact parameters, which are reminiscent of Newman-Janis shifts. For $I_1$ and $\tilde{I}_1$ similar shifts occur, but  are   accompanied by additional contributions that cannot be captured by simply shifting the scalar $I_1$ and $\tilde{I}_1$ waveforms. 
We also show the absence of leading-order corrections to gravitational memory.  
 Our analytic results are  notably compact,  and we compare the effectiveness of the three methods used to obtain them.  We also briefly comment on the magnitude of the corrections to observables due to cubic deformations.

}

\vspace{-2.6cm}

%\pagenumbering{roman}
\maketitle

\flushbottom
 \tableofcontents
\newpage 
%please don't remove this newpage otherwise the intro has no page number

\section{Introduction}

The high accuracy achieved in current observations of gravitational waves by the LIGO-Virgo-KAGRA collaboration, along with the potential of further  improvements  in upcoming experiments such as LISA, strongly motivates the pursuit of 
increasing precision in theoretical calculations.
One is then faced with the need to perform higher-loop calculations in Newton's constant 
$G$, incorporating  spin effects, and also considering    potential modifications to the Einstein-Hilbert theory arising from as yet undiscovered higher-dimensional interactions.

Quadratic corrections in the curvatures are known to leave scattering amplitudes invariant \cite{Tseytlin:1986zz,Deser:1986xr,Tseytlin:1986ti,AccettulliHuber:2019jqo}, hence the first  deformations to be studied appear at dimension six, that is they are cubic in the curvatures (see \cite{Endlich:2017tqa} for work on dimension-eight operators). 
Specifically, we consider the deformations 
\begin{align}
I_1 &\coloneq  {R^{\alpha \beta}}_{\mu \nu} {R^{\mu \nu}}_{\rho \sigma} {R^{\rho \sigma}}_{\alpha \beta}\ , \qquad 
I_2 \coloneq {R^{\mu \nu \alpha}}_\beta {R^{\beta \gamma}}_{\nu \sigma} {R^\sigma}_{\mu \gamma \alpha}\ .  
\end{align}
However,  instead of    $I_2$,  we  prefer working with  the combination 
\begin{align}
    G_3 \coloneq I_1 - 2 I_2\  ,  
\end{align}
which is a topological term in six dimensions and has vanishing four-dimensional graviton amplitudes. 
We will also study the effect of  the parity-odd couplings $\tilde{I}_1$ and $\tilde{G_3}$, which are obtained from the parity-even ones by replacing one of the Riemann curvatures $R^{\mu\nu\alpha\beta}$ by the dual $\tilde R^{\mu\nu\alpha\beta} {=}  
(1/2) \,\epsilon^{\mu\nu\rho\sigma} {R_{\rho\sigma}}^{\alpha\beta}$.
Summarising, the effective action we will work with is 
\begin{equation}
\begin{split}
\label{action}
S = \int\!d^4x \, \sqrt{-g} \,  \bigg(& -\frac{2}{\kappa^2}R   \,  + \, \beta_1 I_1 + \beta_2 G_3 + \tilde{\beta}_1 \tilde{I}_1 + \tilde{\beta}_2 \tilde{G}_3\bigg)\, .
\end{split}
\end{equation}
Particular choices of the coefficients $\beta_1, \beta_2, \tilde{\beta}_1, \tilde{\beta}_2$  correspond to specific theories%
\footnote{For instance $\beta_1 {=} -\frac{2}{\kappa^2} \frac{\, \alpha^{\prime \, 2}}{48}$,  $\beta_2 {=} -\frac{2}{\kappa^2} \frac{\, \alpha^{\prime \, 2}}{24}$, $\tilde{\beta}_{1}{=}\tilde{\beta}_{2}{=}0$ reproduces the low-energy effective action of bosonic string theory.}, but we will be agnostic and treat each deformation independently.  

Cubic  deformations  have been studied  
both using  general relativity approaches \cite{Bueno:2016xff,Hennigar:2017ego,Silva:2022srr}, as well as 
amplitude methods applied to the case of spinless binaries \cite{Brandhuber:2019qpg, Emond:2019crr, AccettulliHuber:2020oou,AccettulliHuber:2020dal}, or systems where only one black hole is spinning and has a  much larger mass than the other black hole  \cite{Burger:2019wkq}.
Related work on other modified theories of gravity include \cite{Sennett:2019bpc,deRham:2019ctd,deRham:2020ejn,deRham:2021bll,CarrilloGonzalez:2022fwg,Melville:2024zjq,Silva:2024ffz,Falkowski:2024bgb}. We also note that in \cite{Silva:2022srr} a bound of $\ell_{\rm EFT}\leq 38.2$~km was determined for the fundamental length scale of cubic theories, specifically for the  case where $\beta_1 {=} \tilde{\beta}_1$, $\beta_2 {=} \tilde{\beta}_2{=}0$. Furthermore,   \cite{Liu:2024atc} investigated the potential  detectability of cubic deformations of gravity in a gravitational-wave event from the merger of two stellar-mass black holes. 

In a typical encounter, we expect both celestial bodies to have non-zero spin vectors, along with arbitrary masses, and this is the scenario we considered in \cite{Brandhuber:2024bnz}. 
In that paper we focused   on  elastic processes, related to the two-to-two scattering amplitude of Kerr black holes,  and   using the KMOC formalism \cite{Kosower:2018adc} we computed the leading-order  impulse (or momentum kick) and spin kick in a generic   hyperbolic encounter. These observables  vanish at tree level, and hence a one-loop calculation was required. 
For the sake of gravitational-wave observations, the most relevant quantity to compute is the waveform, and this is what we  focus on in this paper.

Waveforms for hyperbolic spinning encounters in general relativity were derived to leading order using a worldline quantum field theory approach in \cite{Jakobsen:2021smu} for the spinless case, reproducing the classic results of Kovacs and Thorne \cite{Kovacs:1978eu}. This was later extended to include spinning bodies in   \cite{Jakobsen:2021lvp}, with results valid  up to quadratic order in the spin. Results for the scattering waveform of spinless bodies at next-to-leading order were later derived using amplitude approaches in \cite{Brandhuber:2023hhy, Herderschee:2023fxh,Elkhidir:2023dco, Georgoudis:2023lgf},  while for spinning objects, 
leading-order expressions 
valid to high order in the spins were obtained in \cite{DeAngelis:2023lvf,Brandhuber:2023hhl,Aoude:2023dui}, also using amplitude methods,   which could readily be extended to any spin order once the relevant Compton amplitudes are available.

In order to derive the waveforms  we will follow the adaptation of the KMOC approach presented in \cite{Cristofoli:2021vyo}. In that paper it was shown that, to leading order in the coupling, gravitational  waveforms can be  expressed as a Fourier transform to impact parameter space of the five-point amplitude describing the scattering of two spinning celestial objects with the emission of a graviton. In fact, only the physical singularities of this amplitude contribute to the waveform; as we approach the poles, the five-point amplitude factorises onto a four-point Compton amplitude and a three-point amplitude of two spinning objects with one graviton. 
Compact expressions for  the Compton amplitude in the presence of cubic deformations were derived  in \cite{Brandhuber:2024bnz}, and we will use them to derive all-order in spin expressions for the tree-level waveforms of two Kerr black holes in a cubic background.

After constructing the integrand from factorisation, we are left with the task of taking its Fourier transform. We will do this using three different approaches, which will give us the possibility of comparing their own merits. The first approach  is that of direct integration, building on  the seminal work of \cite{Jakobsen:2021lvp}.
In this method, it is possible to write down ans\"{a}tze for certain tensor (i.e.~not necessarily scalar) integrals which are invariant under rescaling of the integration variable. Such integrals often contain poles on the integration contour and, for the rescaling argument to be valid, they  must  be regularised with a principal value prescription. This is possible  since  additional contributions arising from the   $i \varepsilon$ prescriptions 
 are free of physical poles and hence cancel, as we have checked explicitly.  
 Higher-rank tensor integrals can then be produced efficiently with a generating function technique which is applicable thanks to the presence of a Fourier transform to impact parameter space. Effectively, integrals with  (loop-momentum dependent) numerators can be obtained by differentiating a master integral with a shifted impact parameter with respect to certain scalar auxiliary variables. This approach avoids  differentiation with respect to   four-momenta as well as lengthy Lorentz contractions, thereby leading to very compact results. 

 A second independent approach is based on integration by parts (IBP) reductions and the method of differential equations \cite{Gehrmann:1999as,Kotikov:1990kg,Henn:2013pwa}. The waveforms corresponding to $I_1$ and its parity-odd version $\tilde{I}_1$
contain spurious poles of the form $(\sinh x)/x$, which we avoid by rewriting this function in an integral representation that is well suited for IBP reductions.  As a consequence, the Fourier transforms in the $I_1$ or $\tilde{I}_1$ cases are reduced to a single  simple master integral. We also note that this approach is systematic and  can be extended to one-loop waveforms with or without spin.  Finally, a third approach we  pursued employs Cauchy's residue theorem. This is straightforward to implement and highly  efficient, though it generally produces less compact  analytic results  compared to  the   other two methods, with  direct integration   yielding  the most compact results. 
The three approaches give results in 
perfect~agreement. 

Remarkably, we are able to write the spinning waveforms in terms of  Fourier transforms to impact parameter space where most (and, for the $G_3$ and $\tilde{G}_3$ cases, all)
of the spin dependence is encoded in spin-dependent shifts of the impact parameter. These  shifts are reminiscent of the Newman-Janis shift,  which made an appearance  in amplitude contexts in \cite{Guevara:2018wpp,Arkani-Hamed:2019ymq},  
and more recently in the computation of spinning waveforms in \cite{Brandhuber:2023hhl,Aoude:2023dui}. 
For the $G_3$ and $\tilde{G}_3$ waveforms, we are able to write the spinning waveforms entirely in terms of non-spinning ones (with shifted impact parameter). 

The case of waveforms in the presence of parity-odd deformations is very easy to discuss and in fact does not require new calculations: parity-odd waveforms can be obtained from parity-even ones by simply swapping the ``plus'' and ``cross'' polarisations.  
Investigating parity-odd deformations of gravity  is not merely an academic exercise, given the early 
 indications of parity violation in  the Cosmic Microwave Background and  the  
large-scale structure of galaxies \cite{Minami:2020odp,Diego-Palazuelos:2022dsq,Eskilt:2022wav, Komatsu:2022nvu,Eskilt:2022cff,Philcox:2022hkh,Hou:2022wfj,Philcox:2023ffy}.  

A feature of all the leading waveforms in the presence of cubic deformations is that they do not modify the gravitational memory, which is easily proven using the connnection between the memory and soft theorems \cite{Strominger:2014pwa}.

Our  waveform results can be found in our   
\href{https://github.com/QMULAmplitudes/Cubic-Corrections-to-Spinning-Observables-from-Amplitudes}{{\it Cubic Corrections to Spinning Observables from Amplitudes} GitHub~repository}.  

The rest of the paper is organised as follows. 
In Section~\ref{sec:Compton} we summarise the expressions for the classical Compton amplitudes derived in \cite{Brandhuber:2024bnz}. 
In Section~\ref{sec:integrand} we present the relevant formulae to derive the waveforms from  the factorisation diagrams of the five-point amplitudes. We also construct these factorisations using the classical Compton amplitudes and the classical three-point amplitudes of two spinning particles and one graviton. 
In Section~\ref{sec:directint} we present the direct integration method of the waveform, inspired by the work of \cite{Jakobsen:2021lvp}. The method is illustrated in great detail in the Appendices, while in this section  we use the results of the integrals to derive the  final expressions for the  waveforms with  $G_3$ and $I_1$ deformations. 
In Section~\ref{sec:TIGF} we present our second alternative derivation of the waveforms using a  systematic method proposed in \cite{Chen:2024bpf} for the tensor integral generating functions \cite{Feng:2022hyg}.    
A third derivation of the waveforms is then briefly shown  in 
Section~\ref{sec:residues} using residues, much in line with our previous work \cite{Brandhuber:2023hhl}. Section~\ref{sec:pictures} illustrates the waveforms for various values of the relative velocities of the black holes and their spins, for the $G_3$ and $I_1$ deformations. These waveforms  clearly  show the absence of a  contribution to the gravitational memory, which we   demonstrate  in 
Section~\ref{sec:memory}. 
Finally, in 
Section~\ref{sec:conclusions}  
we conclude by comparing the size of cubic corrections to that of Post-Minkowskian (PM) corrections.

Two Appendices deal with the evaluation of  the  integrals needed to compute our waveforms. Specifically, in 
Appendix~\ref{sec:AppA} we evaluate the necessary master integrals using a combination of explicit evaluations and educated guesses; and in 
Appendix~\ref{sec:AppB} we present the generating function technique described earlier, which allows to compute with great economy all the necessary higher-tensor integrals.

\section{Summary of classical spinning Compton amplitudes with cubic deformations}
\label{sec:Compton}

In this section we briefly review the Compton amplitudes in the presence of parity-even and parity-odd cubic deformations derived in \cite{Brandhuber:2024bnz}. 
We denote by $k_i$ and $\varepsilon_{k_i}$, $i=1,2$, the momenta and polarisations of the emitted gravitons, while $m$ and $a$ are the mass and ring radius of the black hole, with $p$ being the classical momentum.  

\subsection{Parity-even deformations}
For the parity-even cubic deformations, the classical Compton amplitudes 
in the gauge $k_1 \Cdot\varepsilon_{k_2} = k_2 \Cdot\varepsilon_{k_1} =0$ 
are found to be%
\footnote{From now on,   we will omit the coefficients of the cubic interactions $\beta_1, \beta_2, \tilde{\beta}_1, \tilde{\beta_2}$, which  can  be reintroduced at the end of the calculations. Furthermore, we  note that in our normalisations Newton's  constant is defined as   $G \coloneq \kappa^2 /(32\pi)$.} 
\cite{Brandhuber:2024bnz}
\begin{align}
    \begin{split}
    \label{I1amplitude}
    M_{I_1}(p, k_1, k_2) &= i\,  \left( \frac{\kappa}{2}\right)^4   24\,(k_1 \Cdot k_2)^2 (\varepsilon_{k_1}\Cdot 
    \varepsilon_{k_2}) 
    \bigg\{ \cosh ( a\Cdot q ) \bigg[ 2 (p\Cdot \varepsilon_{k_1}) (p\Cdot \varepsilon_{k_2}) - m^2 (\varepsilon_{k_1} \Cdot\varepsilon_{k_2})  \Big]\\
 & -i \frac{\sinh (a\Cdot q)}{a \Cdot q}\Big[ 
(p\Cdot \varepsilon_{k_1} ) \epsilon( \varepsilon_{k_2} p a q) +(p\Cdot \varepsilon_{k_2} ) \epsilon( \varepsilon_{k_1} p a q) 
\Big] \bigg\} \, , 
  \end{split}
    \end{align}
    and
\begin{align}
    \begin{split}
    \label{I2amplitude}
    M_{I_2}(p, k_1, k_2) &=
    i  \,  \left( \frac{\kappa}{2}\right)^4 6 \, (k_1 \Cdot k_2)  (\varepsilon_{k_1}\Cdot 
    \varepsilon_{k_2})^2  
    \bigg\{\cosh ( a\Cdot q ) \Big[ \big( p\Cdot (k_1 - k_2)\big)^2 + m^2 (k_1 \Cdot k_2)   \Big]\\
 & -2i \, \frac{\sinh (a\Cdot q)}{a \Cdot q} p\Cdot (k_1 - k_2)\  \epsilon( k_1 k_2  p a)  
\Big] \bigg\} \, , 
  \end{split}
    \end{align}
    while 
    \begin{align}
        M_{I_1}(p, k_1^{\pm\pm}, k_2^{\mp\mp})= M_{I_2}(p, k_1^{\pm\pm}, k_2^{\mp\mp})=0 
        \ ,  
        \end{align}
        where 
        $q=k_1 + k_2$.
        
The amplitudes  \eqref{I1amplitude} and \eqref{I2amplitude} can also be rewritten in spinor-helicity variables:%
\footnote{Our spinor conventions are the same as in  \cite{Brandhuber:2022qbk}.} 
\begin{align}
\begin{split}
\label{I1amplitude-bis}
    M_{I_1}(p, k_1^{++}, k_2^{++}) &= 
    i\left( \frac{\kappa}{2}\right)^4 3! \frac{[1\, 2]^4} {q^2}\bigg\{ - 4 \cosh(a\Cdot q)  \ (p\Cdot k_1)(p\Cdot k_2)\,    \\
 & +\frac{1}{2}p\Cdot(k_1-k_2) \frac{\sinh (a\Cdot q)}{a \Cdot q}\Big( [1|p|2\rangle [2|a|1\rangle - [2|p|1\rangle [1|a|2\rangle \Big) 
 \bigg\} \, , 
  \end{split}
    \end{align}
    and 
    \begin{align}
    \label{I2amplitude-bis}
        \begin{split}
            M_{I_2}(p, k_1^{++}, k_2^{++}) &= \frac{i}{2} \left( \frac{\kappa}{2}\right)^4 \, 3!\, \frac{[1\, 2]^4}{q^2}\bigg\{ \cosh (a\Cdot q) \Big[ \big( p\Cdot (k_1 - k_2)\big)^2 + m^2 (k_1 \Cdot k_2)   \Big]
            \\ 
            &+\frac{1}{2}p\Cdot(k_1-k_2) \frac{\sinh (a\Cdot q)}{a \Cdot q}\Big( [1|p|2\rangle [2|a|1\rangle - [2|p|1\rangle [1|a|2\rangle \Big)  \bigg\}\ . 
        \end{split}
    \end{align}
In the following we will focus on the $G_3$ deformations instead of $I_2$. The corresponding classical Compton amplitude in spinor-helicity variables is  
\begin{align}
\label{G3amplitude}
    M_{G_3}(p, k_1^{++}, k_2^{++}) = -3 \, i\,m^2  \left( \frac{\kappa}{2}\right)^4 \cosh (a\Cdot q) [1\, 2]^4 \, 
    .
\end{align}
Note that this amplitude has no term proportional to $(\sinh a\Cdot q) /(a\Cdot q)$.

\subsection{Parity-odd deformations}

Finally, as discussed in \cite{Brandhuber:2024bnz}
the amplitudes with parity-odd deformations can be obtained easily from the parity-even  ones:  for the case of positive (negative)  helicity, they are equal to those for the corresponding parity-even case multiplied by a factor of  $+i$ ($-i$), and we recall that the only non-vanishing amplitudes are those where the two gravitons have the same helicity.

\section{Constructing the time-domain waveforms}
\label{sec:integrand}

\subsection{General results}

The derivation of the time-domain waveforms using the KMOC formalism was presented  in \cite{Cristofoli:2021vyo}.  
We will work in the  far-field limit, that is at large observer distance $r{\coloneqq }|\vec{x}|$  and time $t$ with fixed retarded time $u{\coloneqq} t{-}r$.  
The quantity of interest  is then 
\begin{align}
    \begin{split}
    \langle h_{+}\pm i h_{\times}\rangle &\coloneqq \langle h_{\mu \nu}^{\rm out}\rangle  \vareps^{\mu \nu}_{(\pm \pm)}
    \coloneqq\frac{1}{4\pi r }(h_{+}^\infty\pm i h_{\times}^\infty) \, ,  
    \end{split}
\end{align}
for which one finds that%
\footnote{See for example Section~5 of \cite{Brandhuber:2023hhl} for a detailed derivation.} 
\begin{align}
\label{KMOCsubfinalforhmunuquater}
	\begin{split}
		& h^{\infty}_{+} \pm i  h^{\infty}_{\times}  =    
  %\frac{\kappa}{4\pi r} 
  \kappa    
	   \int_{-\infty}^{+\infty}\!\frac{d\omega}{2\pi} e^{-i \omega u } %\\ &
 \Big[ 	 
		 %\varepsilon^{(h)\ast}_{\mu\nu}(\hat{k}) \, 
   \theta(\omega)\,  \left.{W}\big(b, k^\pm\big)  \right|_{k= \omega (1, \mathbf{\hat{x}})}  
  %\\ 
 +  \, 
 %\varepsilon^{(h)}_{\mu\nu}(\hat{k})
 \theta(-\omega)\,  \left. {W}^\ast\big(b,k^\mp\big) \right|_{k= -\omega (1, \mathbf{\hat{x}})}\Big]
		, 
	\end{split}
\end{align}
where  ${W}{=}{W}(\vec{b}, k; h)$  is the spectral waveform for the emission of a graviton of momentum $k$ and helicity $h$, 
which satisfies,  at tree level, 
    \begin{align}
\label{seconda}
\left.    W^\ast (b, k^h)\right|_{k=-\omega(1, \hat{\mathbf{x}})} = 
\left.    W(b, k^{-h})\right|_{k=\omega(1, \hat{\mathbf{x}})}\, . 
\end{align}
Using this result, one can easily combine  the two terms in \eqref{KMOCsubfinalforhmunuquater} into 
\begin{align}
\label{combinedres}
   h^{\infty}_{+} \pm i  h^{\infty}_{\times} = 
\kappa \int_{-\infty}^{+\infty}\!
\frac{d\omega}{2\pi}e^{- i \omega u} 
 \left. 
W(b, k^{\pm})\right|_{k=\omega(1, \hat{\mathbf{x}})}
 \, . 
\end{align}
At leading order (tree level) this can be neatly obtained from the  five-point classical bremsstrahlung amplitude $\mathcal{M}_{5}(q_1, q_2, a_1, a_2; h)$ describing the scattering of two objects with ring radii $a_1$ and $a_2$ with the emission of a graviton with momentum $k{=} q_1 + q_2$ and helicity $h$: 
\begin{align}
\label{KMOCsubfinalbis}
{W}(b, k^h ) \coloneqq  -i \int\!d\mu^{(4)}\  e^{i(q_1\Cdot b_1 + i q_2 \Cdot b_2) } \ \mathcal{M}_{5}(q_1, q_2, a_1, a_2; h) \, , 
\end{align}
where  
\begin{align}
\label{measureinnn}
	d\mu^{(D)} \coloneqq \frac{d^Dq_1}{(2\pi)^{D{-}1}} \frac{d^Dq_2}{(2\pi)^{D{-}1}}
	\, (2\pi)^D  \delta^{(D)} (q_1 + q_2 - k) \delta(2  {\pb}_1\Cdot q_1 ) \delta(2  {\pb}_2\Cdot q_2 )\, ,
\end{align}
and   $q_{1,2}{=} p_{1,2} {-} p_{1,2}^\prime$ are the momentum transfers. The barred variables 
\cite{Landshoff:1969yyn,Parra-Martinez:2020dzs}  are defined as 
\begin{align}
	\label{barredv}
	\begin{split}
		p_1 &= \bar{p}_1 + \frac{q_1}{2}\, , \qquad p_1^\prime = \bar{p}_1 - \frac{q_1}{2}\, , \\
		p_2 &= \bar{p}_2 +  \frac{q_2}{2}\, , \qquad p_2^\prime = \bar{p}_2 - \frac{q_2}{2}\, , 
	\end{split}
\end{align}
and satisfy  \begin{align}
	\label{eq: HEFTfame}
	\bar{p}_1\Cdot q_1 =\bar{p}_2\Cdot q_2 =0 \, .  
\end{align}
However, at leading order we can actually drop the distinction between $\bar{p}_i$ and $p_i$, hence we will henceforth drop bars. Furthermore, as in \cite{Brandhuber:2023hhl},
in order to simplify the notation  we will  work with the quantity 
\begin{align}
\begin{split}
\label{KMOCsubfinalbis2}
 h^\infty(u) &\coloneqq 
 \kappa \int_{-\infty}^{+\infty}\!
\frac{d\omega}{2\pi}e^{- i \omega u} 
 \left. 
W(b, k)\right|_{k=\omega(1, \hat{\mathbf{x}})}\\ 
&= -i\kappa \int_{-\infty}^{+\infty}\!
\frac{d\omega}{2\pi}e^{- i \omega u} 
   \int\!  \frac{d^4q_1}{(2\pi)^{2}}  \delta(2  {p}_1\Cdot q_1 ) \delta(2  {p}_2\Cdot (k-q_1) )\  e^{i(q_1\Cdot b_1 + q_2 \Cdot b_2)} \ \mathcal{M}_5 \, ,
\end{split}
\end{align}
where we omitted the helicity dependence, 
performed the $q_2$ integration using momentum conservation from \eqref{measureinnn}
and defined
\begin{align}
\label{khat}
k\coloneq \omega \hat{k} =\omega(1, \hat{\mathbf{x}})\, . 
\end{align}
This is the main formula we will use. We also note that $b{\coloneq} b_{1} - b_{2}$ is the asymptotic impact parameter, which  satisfies 
\begin{align}
b\Cdot v_1 = b\Cdot v_2 = 0\, . 
\end{align}

\subsection{Kinematics of the five-point scattering}

The time-domain waveform \eqref{KMOCsubfinalbis2} requires knowledge of the classical part of the five-point bremsstrahlung amplitude, that is the amplitude of two spinning particles with  one emitted graviton. 
The kinematics of the process is shown below: 
\begin{equation}\label{eq: kinematics}
	\begin{array}{lr}

		\begin{tikzpicture}[scale=12,baseline={([yshift=-1mm]centro.base)}]
			\def\x{0}
			\def\y{0}

			\node at (0+\x,0+\y) (centro) {};
			\node at (-3pt+\x,-3pt+\y) (uno) {\footnotesize $p_1$};
			\node at (-3pt+\x,3pt+\y) (due) {\footnotesize $p_2$};
			\node at (3pt+\x,3pt+\y) (tre) {\footnotesize $p_2^\prime $};
			\node at (3pt+\x,-3pt+\y) (quattro) {\footnotesize $\ \ p_1^\prime $};
			\node at (5.55pt+\x,\y) (cinque)
			% era 4.25 
			{\footnotesize $k= q_1 + q_2$};

			\draw [thick,double,orange] (uno) -- (centro);
			\draw [thick,double,amber] (due) -- (centro);
			\draw [thick,double,amber] (tre) -- (centro);
			\draw [thick,double,orange] (quattro) -- (centro);
			\draw [vector,double] (cinque) -- (centro);

			\draw [->] (-2.8pt+\x,-2pt+\y) -- (-1.8pt+\x,-1pt+\y);
			\draw [<-] (2.8pt+\x,-2pt+\y) -- (1.8pt+\x,-1pt+\y);
			\draw [<-] (-1.8pt+\x,1pt+\y) -- (-2.8pt+\x,2pt+\y);
			\draw [->] (1.8pt+\x,1pt+\y) -- (2.8pt+\x,2pt+\y);
			\draw [->] (2.3pt+\x,-0.6pt+\y) -- (3.75pt+\x,-0.6pt+\y);

			\shade [shading=radial] (centro) circle (1.5pt);
		\end{tikzpicture}
		 & \hspace{2cm}
		%	\begin{aligned}
		%& k^\mu = q_1^\mu + q_2^\mu\ ,     \\
		% & p_i^\mu =  m_i v_i^\mu 
		%\ ,       \\
		%& p_i \Cdot q_i \sim 0 
		%  \ .
		%	\end{aligned}
	\end{array}
\end{equation}
The  two Kerr black holes  have  masses $m_1$ and $m_2$  with 
\begin{align}
 p_1^2 = (p_1^\prime)^2 = m_1^2\, , \qquad  p_2^2 = (p_2^\prime)^2 = m_2^2\, . 
\end{align}
It is also convenient to  introduce four-velocities $v_1$ and $v_2$ as
\begin{align}
    p_1 \coloneq m_1 v_1\, , \qquad  p_2 \coloneq m_2 v_2\, , 
\end{align}
so that  $v_1^2 = v_2^2 = 1$. The two black holes have ring radii $a_1$ and $a_2$, where for a single black hole we recall that the ring radius $a^\mu$ is related to the spin vector as $a^\mu \coloneq S^\mu/ m$~\cite{Vines:2017hyw,Guevara:2018wpp,Chung:2019yfs,Bern:2020buy,Bautista:2021wfy}, where 
\begin{align}
\label{def:spinvector}
S^\mu = \frac{1}{2 m} \epsilon^{\mu\nu\alpha\beta} p_{\nu} S_{\alpha\beta} \ , 
\end{align}
with $S^{\mu \nu}$ being the spin tensor. 
The ring radius satisfies the covariant spin supplementary condition    $p\Cdot a(p) =0$. 
 Finally, we also define the Lorentz factor
\begin{equation}
	\label{fiveiv}
	\begin{split}
		\sigma \coloneqq v_1 \Cdot v_2  \geq 1\ .
	\end{split}
\end{equation}
Note that $\sigma=\frac{1}{\sqrt{1-\dot{\vec{x}}^2}}$, with  $\dot{\vec{x}}$ being the relative velocity of one of the two black holes  in the rest frame of the other. For instance, in  the rest frame of particle $1$ we can write  $v_1^{\mu}=(1,0,0,0)$ and   $v_2^{\mu}=\sigma(1, \dot{\vec{x}})$. 

\subsection{Waveform integrands}

An important simplification  in the calculation of the waveforms  consists in the fact that only the residues on the physical factorisation channels  are needed in order to obtain  the waveform. These can  be computed from the two factorisation diagrams 
\begin{align}
\label{2factordiagrams}
    \begin{tikzpicture}[baseline={([yshift=-0.4ex]current bounding box.center)}]\tikzstyle{every node}=[font=\small]
		\begin{feynman}
			\vertex (p1) {\footnotesize $1$};
			\vertex [above=1.5cm of p1](p2){\footnotesize $2$};
			\vertex [right=1.5cm of p2] (u1) [HV]{H};
			\vertex [right=1.2cm of u1] (p3){\footnotesize $2^\prime$};
			\vertex [right=1.5cm of p1] (b1) [dot]{};
			\vertex [right=1.2cm of b1] (p4) []{\footnotesize $1^\prime$};
			\vertex [above=0.4cm of p1] (cutLt);
            \vertex [right=0.6cm of cutLt] (cutL);
			\vertex [right=1.8cm of cutL] (cutR){\footnotesize $k$};
			\vertex [right=0.5cm of b1] (cut1);
			\vertex [above=0.2cm of cut1] (cut1u);
			\vertex [below=0.2cm of cut1] (cut1b);
            \draw[vector, double] (u1) -- (b1); % so the cut line is drawn on top of the graviton
			\diagram* {
			(p2) -- [thick, amber, double] (u1) -- [thick, double, amber] (p3),
			(b1)--[opacity=0,momentum=\footnotesize$q_1$](u1),  (u1)-- [vector, double] (cutR), (p1) -- [thick, orange, double] (b1)-- [thick, orange, double] (p4), (cutL)--[dashed, red,thick] (cutR),
			};
		\end{feynman}
	\end{tikzpicture}  &&
\begin{tikzpicture}[baseline={([yshift=-0.4ex]current bounding box.center)}]\tikzstyle{every node}=[font=\small]
		\begin{feynman}
			\vertex (p1) {1};
			\vertex [above=1.5cm of p1](p2){\footnotesize $2$};
			\vertex [right=1.5cm of p2] (u1) [dot]{};
			\vertex [right=1.2cm of u1] (p3){\footnotesize $     2^\prime$};
			\vertex [right=1.5cm of p1] (b1) [HV]{H};
			\vertex [right=1.2cm of b1] (p4) []{\footnotesize $1^\prime$};
			\vertex [above=1.1cm of p1] (cutLt);
            \vertex [right=0.6cm of cutLt] (cutL);
			\vertex [right=1.8cm of cutL] (cutR){\footnotesize $k$};
			\vertex [right=0.5cm of b1] (cut1);
			\vertex [above=0.2cm of cut1] (cut1u);
			\vertex [below=0.2cm of cut1] (cut1b);
            \draw[vector, double] (b1) -- (u1); % so the cut line is drawn on top of the graviton
			\diagram* {
			(p2) -- [thick, amber, double] (u1) -- [thick, amber, double] (p3),
			(u1)--[opacity=0,momentum'=\footnotesize$q_2$](b1), (b1)-- [vector, double] (cutR), (p1) -- [thick, orange, double] (b1)-- [thick, double, orange] (p4), (cutL)--[dashed, red,thick] (cutR),
			};
		\end{feynman}
	\end{tikzpicture} 
 \end{align}
which  correspond to the two possible factorisations as $q_1^2{\to} 0$ or $q_2^2{\to} 0$, with $q_1 + q_2 {=} k$. 
In each diagram, the key ingredients are the Compton amplitudes for two spinning objects, which were computed for the parity-even and parity-odd cubic deformations  in~\cite{Brandhuber:2024bnz} and are quoted in Section~\ref{sec:Compton}. Using these, we now construct the integrands for the  parity-even and parity-odd cubic deformations of gravity. 

For simplicity, from now on we focus on the case of an emitted graviton with positive helicity. The two factorisation diagrams are then given by 
\begin{align}
\label{2factordiagrams2}
\cM_{I, q_1^2} & = \frac{i}{q_1^2} M_3(-p_1, q_1^{--}) M_I(-p_2, - q_1^{++}, k^{++})  \, , \\
\cM_{I, q_2^2} & = \frac{i}{q_2^2} M_3(-p_2, q_2^{--}) M_I(-p_1, - q_2^{++}, k^{++}) \, , 
\end{align}
with  
\begin{align}
\label{M5fact}
\cM_{5, I}  = \cM_{I, q_1^2}+ \cM_{I, q_1^2} + {\rm contact \ terms}
\, . 
\end{align}
Here  $M_I$ denotes  the Compton amplitude in the presence of a deformation $I$,  with $I{\in} (I_1, G_3, \tilde{I}_1, \tilde{G}_3)$, as given in Section~\ref{sec:Compton}. 
$M_3$ is  
a classical three-point amplitude with one graviton of momentum $q$ and two massive 
spinning particles with momenta $p$ and $-(p+q)$ and ring radius~$a$. For the two possible helicities of the emitted gravitons one has~\cite{Guevara:2018wpp}
\begin{align}
\label{3pc}
    M_3 (p, q^{++}) = -i \left( \frac{\kappa}{2}\right) e^{a\Cdot q} \left(
    \frac{\langle \xi | p | q]}{\langle \xi q\rangle }
    \right)^2\, , \qquad 
    M_3 (p, q^{--}) = -i \left( \frac{\kappa}{2}\right) e^{-a\Cdot q} \left(
    \frac{\langle q  | p | \xi ]}{[ q \xi] }
    \right)^2\, , \qquad 
\end{align}
where $\xi$ and $\tilde{\xi}$ are reference spinors. 
Finally, we observe that the second factorisation diagram in \eqref{2factordiagrams} can be obtained from the first as 
\begin{align}
    \cM_{I, q_2^2} = \left.\cM_{I, q_1^2}\right|_{1\leftrightarrow2}\, ,  
\end{align}
where $1\leftrightarrow2$ means swapping  the masses, spin vectors and momenta, in addition to also exchanging $q_1$ and $q_2$.

With these ingredients it is very easy to construct the integrand of the waveforms. For the case of the $G_3$ deformation,  and choosing as reference spinor of the three-point amplitude the spinor of the external graviton, we get 
\begin{align}
\label{firstG3iint}
   \cM_{G_3, q_1^2} = - i \left(\frac{\kappa}{2}\right)^5\frac{3  m_1^2 m_2^2}{q_1^2}  [k| v_1 q_1 | k]^2\, e^{- a_1 \Cdot q_1}\cosh\big( a_2\Cdot (q_1 - k) \big) \ .   
\end{align}
For $I_1$, the result is
\begin{align}
\begin{split}
   \cM_{I_1, q_1^2} &=  -i \left(\frac{\kappa}{2}\right)^5\frac{3  m_1^2 m_2^2}{q_1^2}  
  \frac{[k| v_1 q_1 | k]^2}{ q_1\Cdot k}e^{- a_1 \Cdot q_1} \times
  \\ \!\!\!\!& \!\!\!\!\bigg[ - 4 \cosh [ a_2\Cdot (k-q_1)] (-v_2\Cdot q_1) (v_2 \Cdot k) - 2 i  v_2 \Cdot (k+q_1) \frac{\sinh a_2\Cdot (k-q_1)}{a_2\Cdot (k-q_1)}\epsilon ( k v_2 q_1 a_2)\bigg] \ . 
  \end{split}
\end{align}
Note that because of the first  $\delta$-function in \eqref{KMOCsubfinalbis2}
we can simplify the above expression using $v_1\Cdot q_1{=}0$ into 
\begin{align}
\begin{split}
\label{firstI1iint}
   \cM_{I_1, q_1^2} &= -  i \left(\frac{\kappa}{2}\right)^5\frac{12\,   m_1^2 m_2^2}{q_1^2\, (q_1\Cdot k)}  \, 
  [k| v_1 q_1 | k]^2(v_2 \Cdot k )\, e^{- a_1 \Cdot q_1} \times
  \\ & \bigg[   \cosh [ a_2\Cdot (k-q_1)]  (v_2 \Cdot k) -  i\,    \frac{\sinh a_2\Cdot (k-q_1)}{a_2\Cdot (k-q_1)}\epsilon ( k v_2 q_1 a_2)\bigg] \ . 
  \end{split}
\end{align}

\section{Time-domain waveforms from direct integration }
\label{sec:directint}

There are several approaches to explicitly perform the integrations in \eqref{KMOCsubfinalbis2}. 
In this section  we follow  the direct integration approach of \cite{Jakobsen:2021smu,Jakobsen:2021lvp}, while in Section~\ref{sec:residues} we present  an alternative derivation  based on Cauchy's  residue theorem.

The first step in the direct integration method consists in 
performing the $\omega$ integration
in \eqref{KMOCsubfinalbis2} using
the delta function $\delta (2 p_2 \Cdot (k-q_1))$. This localises the graviton energy $\omega$ to the value $\omega^\ast$ given by  
\begin{align}
\label{omast}
\omega^\ast = \frac{q_1\Cdot v_2}{\hat{k}\Cdot v_2}\, , 
\end{align}
where we recall from \eqref{khat} that 
   $k\coloneq \omega \hat{k}$.
One then quickly arrives at \cite{Jakobsen:2021lvp}
\begin{align}
\label{inttodo}
    h^\infty(u) = - \frac{i \kappa}{2 m_1 m_2 (\hat{k}\Cdot v_2)}
    \int\!\frac{d^4 q_1}{(2\pi)^3}\delta(2 q_1 \Cdot v_1) e^{i q_1\Cdot \tilde{b}} \left.\cM_{5}\right|_{k = \omega^\ast \hat{k}}\, , 
\end{align}
with the following definitions: 
\begin{align}
\begin{split}
\label{btildeetal}
    \tilde{b} &\coloneq \tilde{b}_1 - \tilde{b}_2\, ,  
    \\ \tilde{b}_i & \coloneq b_i + u_i v_i\, , \qquad 
   u_i  \coloneq \frac{u - \hat{k}\Cdot b_i}{\hat{k}\Cdot v_i}\, , \qquad i=1,2. 
\end{split}
\end{align}
We note that the modified (shifted) impact parameter $\tilde{b}$ has the useful  property that 
\begin{align}
    \hat{k}\Cdot \tilde{b}=0\, , 
\end{align}
since $\hat{k}\Cdot \tilde{b}_i {=}u$. We are then left to perform integrations of the form \eqref{inttodo}. For clarity we have derived and collected all required  master integrals in Appendix~\ref{sec:AppA} 
and~\ref{sec:AppB}, where a detailed description of the direct integration method is included.

\subsection{The $G_3$ waveform}
We now discuss the $G_3$ waveform. 
The relevant integral is \eqref{inttodo}, where we can replace the five-point amplitude with the contributions of the two poles,  $\cM_{G_3, q_1^2}$, $\cM_{G_3, q_2^2}$ and discarding contact terms (see \eqref{M5fact}).   
To proceed, we first rewrite  
\begin{align}
    |k] \coloneq \sqrt{\omega} |\hat{k}]\, , 
\end{align}
and recall that $\omega$ is localised at the $q_1$-dependent value $\omega^\ast$ given in \eqref{omast}, and where $| \hat{k}]$ is $\omega$-independent. 
We also rewrite the $\cosh$ term in $M_{G_3}(-p_2, - q_1^{++}, k^{++})$ in terms of exponentials, with  the net effect of shifting the impact parameters. Doing so, we obtain for  
the contribution to the time-domain waveform  arising from the first diagram in \eqref{2factordiagrams}, 
\begin{align}
\begin{split}
\label{interG3}
    &h^{\infty, (1)}_{G_3} (u) = - \frac{i \kappa}{2 m_1 m_2 (\hat{k}\Cdot v_2)}
    \int\!\frac{d^4 q_1}{(2\pi)^3}\delta(2 q_1 \Cdot v_1) e^{i q_1\Cdot \tilde{b}} \left.\cM_{G_3, q_1^2}\right|_{k = \omega^\ast \hat{k}}
    \\ &=- \frac{3}{2} \left(\frac{\kappa}{2}\right)^6
    \frac{m_1 m_2}{(\hat{k}\Cdot v_2)^3} \int\!\frac{d^4q_1}{(2\pi)^3} \delta(2 q_1\Cdot v_1) \frac{(q_1 \Cdot X)^2 (q_1\Cdot v_2)^2}{q_1^2} \left[ e^{i q_1 \Cdot [  \tilde{b} +  i (\tilde{a}_1 - \tilde{a}_2) ]} + e^{i q_1 \Cdot [  \tilde{b} + i (\tilde{a}_1 + \tilde{a}_2)]}\right] \\ 
    & = - \frac{3}{2}  \left(\frac{\kappa}{2}\right)^6
    \frac{m_1 m_2
    }{(\hat{k}\Cdot v_2)^3} X_\mu X_\nu v_{2 \alpha} v_{2\beta} \Big[ 
    \left.\cI^{\mu \nu \alpha \beta}\right|_{\tilde{b}\to \tilde{b} + i(\tilde{a}_1 - \tilde{a}_2)} + \left.\cI^{\mu \nu \alpha \beta}\right|_{\tilde{b}\to \tilde{b}+ i(\tilde{a}_1 + \tilde{a}_2)}\Big] \\
    & = - \frac{3}{2} \left(\frac{\kappa}{2}\right)^6
    \frac{m_1 m_2
    }{(\hat{k}\Cdot v_2)^3}\Big[ 
    \left.\cC^{(1)}\right|_{\tilde{b}\to \tilde{b} + i(\tilde{a}_1 - \tilde{a}_2)} + \left.\cC^{(1)}\right|_{\tilde{b}\to \tilde{b}+ i(\tilde{a}_1 + \tilde{a}_2)}\Big] \ , 
\end{split}
\end{align}
where we used the expression for $\cM_{G_3, q_1^2}$ derived earlier in \eqref{firstG3iint}. We also  defined
\begin{align}
X_\mu = [\hat{k}| v_1 \sigma_\mu |\hat{k}] \, ,  
\end{align}
and
\begin{align}
\label{tildeai}
\tilde{a}_i  \coloneq a_i - v_i \frac{a_i\Cdot \hat{k}}{v_i\Cdot \hat{k}} \ , 
\end{align}
which allows us, using \eqref{omast}, to rewrite%
\footnote{In writing \eqref{interG3} we have also used that $q_1\Cdot v_1{=}0$ and replaced $a_1\Cdot q_1$ with $\tilde{a}_1\Cdot q_1$. This is useful in order to maintain the symmetry between the two factorisation diagrams in \eqref{2factordiagrams} so that the second can easily be obtained from the first upon performing suitable replacements, see \eqref{swap}. 
}
\begin{align}
a_2 \Cdot (k-q_1) = -\tilde{a}_2 \Cdot q_1 \ .
\end{align}
The definitions of the tensor integrals $\cI^{\mu_1 \ldots \mu_n}$ can be found in \eqref{intoneden}. 
 
The integrals in the second line of \eqref{interG3} can be computed elegantly using the method of generating functions introduced in Appendix~\ref{sec:AppB}. In particular  \eqref{G3int} and \eqref{G3intderiv} are relevant for this case, with the result
\begin{align}\label{G3genfun}
\begin{split}
\cC^{(1)}&=\int\!\frac{d^4q_1}{(2\pi)^3} \delta(2 q_1\Cdot v_1) \frac{(q_1 \Cdot X)^2 (q_1\Cdot v_2)^2}{q_1^2} e^{i q_1 \Cdot \tilde{b}} \\
&= -\frac{1}{8 \pi} \left(\frac{\partial^4}{\partial^2 t_1 \partial^2 t_2} \frac{1}{|\tilde{b}_{(1)}+t_1 v_{2(1)} + t_2 X_{(1)}|}\right)_{t_1=t_2=0} \ . 
\end{split}
\end{align}
Here the subscript $(1)$  indicates the projection imposed by the delta function and implemented by the projector 
\begin{align}
\label{P1}
P_1^{\mu\nu}=\eta^{\mu\nu}-v_1^\mu v_1^\nu \ ,
\end{align}
which allows us to define
\begin{align}
\label{V(1)}
V^\mu_{(1)} = P_1^{\mu\nu} V_\nu \ , 
\end{align}
and
\begin{align}
\label{modV(1)}
|V_{(1)}| = \sqrt{-V_{(1)} \Cdot V_{(1)}}
= \sqrt{-V \Cdot P_1 \Cdot V} \ .
\end{align}
We also note that $X^\mu_{(1)} = X^\mu$.

With these definitions and performing the derivatives in \eqref{G3genfun} we find
\begin{align}
\begin{split}
\label{littlevomit-0}
\cC^{(1)} &  = -\frac{3}{8 \pi|\tilde{b}_{(1)}|^5}\bigg[ 
2 (X \Cdot v_{2 (1)})^2\\ & 
+ \frac{5}{|\tilde{b}_{(1)}|^2} \bigg(v_{2(1)}^2 (X \Cdot \tilde{b}_{(1)})^2 
+ 4 (X \Cdot v_{2 (1)}) 
(X \Cdot \tilde{b}_{(1)})(\tilde{b}_{(1)} \Cdot v_{2(1)})
+7 \frac{(\tilde{b}_{(1)}\Cdot v_{2})^2}{|\tilde{b}_{(1)}|^2} (X \Cdot \tilde{b}_{(1)})^2\bigg)
\bigg]\, , 
\end{split}
\end{align}
which can be simplified further into%
\footnote{We have also rewritten $\tilde{b}_{(1)} \Cdot v_{2(1)}$ as $\tilde{b}_{(1)} \Cdot v_{2}$, since for any two vectors $m^\mu$ and $n^\mu$ we have $m_{(1)}\Cdot n_{(1)} =m\Cdot n_{(1)} = m_{(1)}\Cdot n$, with $m_{(1)}^\mu \coloneq P_1^{\mu \nu} m_\nu $, with the projector $P_1$ defined in \eqref{P1}.}
\begin{align}
\label{littlevomit}
\cC^{(1)} = -\frac{3}{8 \pi}\bigg[ 
\frac{2 (X \Cdot v_{2})^2}{|\tilde{b}_{(1)}|^5}  
+ \frac{5 (X \Cdot \tilde{b})}{|\tilde{b}_{(1)}|^7} \bigg(v_{2(1)}^2 (X \Cdot \tilde{b}) 
+ 4 
(\tilde{b}_{(1)} \Cdot v_{2}) (X \Cdot v_{2}) 
+7 \frac{(\tilde{b}_{(1)}\Cdot v_{2})^2}{|\tilde{b}_{(1)}|^2} (X \Cdot \tilde{b})\bigg)
\bigg]\, .
\end{align}
Alternatively, we can evaluate the third line of \eqref{interG3} by twice differentiating \eqref{Imunusimple}, which gives  
\begin{align}
\begin{split}
\label{fourtensorint}
    \cI^{\mu \nu \alpha \beta}(\tilde{b}) & = - \frac{3}{8 \pi|\tilde{b}_{(1)}|^5}\bigg[ 
    \Big( P_1^{\mu \nu} P_1^{\alpha \beta} + P_1^{\mu \alpha} P_1^{\nu \beta} + P_1^{\nu \alpha} P_1^{\mu \beta}  
    \Big) 
    \\
    &+ \frac{5}{|\tilde{b}_{(1)}|^2  }\Big(
    P_1^{\mu \nu}\tilde{b}_{(1)}^{\alpha} \tilde{b}_{(1)}^{ \beta} + P_1^{\mu \alpha} \tilde{b}_{(1)}^{\nu} \tilde{b}_{(1)}^{\beta} + P_1^{\nu \alpha} \tilde{b}_{(1)}^{\mu} \tilde{b}_{(1)}^{\beta}
    \\ & +
    P_1^{\alpha \beta} \tilde{b}_{(1)}^\mu \tilde{b}_{(1)}^\nu + P_1^{\mu \beta} \tilde{b}_{(1)}^\alpha \tilde{b}_{(1)}^\nu + P_1^{\nu \beta} \tilde{b}_{(1)}^\alpha \tilde{b}_{(1)}^\mu \Big)
    \\ & + \frac{35}{|\tilde{b}_{(1)}|^4  } \ \tilde{b}_{(1)}^\mu \tilde{b}_{(1)}^\nu \tilde{b}_{(1)}^\alpha \tilde{b}_{(1)}^\beta\bigg] \, ,      
    \end{split}
\end{align}
where the projector $P_1^{\mu \nu}$ is defined in \eqref{P1}, 
%$\tilde{b}_{(1)}^\mu \coloneq P_1^{\mu \nu} \tilde{b}_\nu$ and 
$\tilde{b}_{(1)}^\mu$ in \eqref{V(1)} and 
$|\tilde{b}_{(1)}|$ 
in \eqref{modV(1)}. 
%we define $|\tilde{b}| \coloneq \sqrt{ - \tilde{b}\Cdot %\tilde{b}}$. 
Note that all the terms proportional to $P_1^{\mu \nu}$ in 
\eqref{fourtensorint} do not contribute to  the contractions in \eqref{interG3}.
Pleasingly, we find complete agreement of the contractions of this tensor integral with the result obtained from the generating function,  
\begin{align}
\label{littlevom}
X_\mu X_\nu v_{2 \alpha} v_{2\beta}\ \cI^{\mu \nu \alpha \beta}(\tilde{b}) = \cC^{(1)} \ .
\end{align}
We then have 
\begin{align}
\begin{split}
\label{G3-1}
    h^{\infty, (1)}_{G_3} (u)  
    & = - \frac{3}{2}   \left(\frac{\kappa}{2}\right)^6
    \frac{m_1 m_2
    }{(\hat{k}\Cdot v_2)^3}\Big[ 
    \left.\cC^{(1)}\right|_{\tilde{b}\to \tilde{b} + i(\tilde{a}_1 - \tilde{a}_2)} + \left.\cC^{(1)}\right|_{\tilde{b}\to \tilde{b}+ i(\tilde{a}_1 + \tilde{a}_2)}\Big] 
    \, . 
\end{split}
\end{align}
We  note that effectively we have rewritten the spinning waveform in terms of the scalar waveform, but with shifted impact parameters: 
\begin{align}
  h_{G_3}^{\infty, (1)} (u) =\frac{1}{2} \left[ \left. h_{G_3}^{\infty, (1)} (u) \right|_{a_i=0, \ \tilde{b}\to \tilde{b}+{ i (\tilde{a}_1 + \tilde{a}_2)}} 
\, + \,  \left. h_{G_3}^{\infty, (1)} (u) \right|_{a_i=0, \ \tilde{b}\to \tilde{b}+{ i (\tilde{a}_1 - \tilde{a}_2)}}\right] \, .   \end{align} 
We call the two contributions of these two terms as arising from two ``sectors'', with shifted impact parameter
\begin{align}
    {\bf (1)}~~~~~~~~&& {\bf (2)}~~~~~~~~\nn\\
   \hat{b} = \tilde b + i (\tilde a_1+ \tilde a_2)  &&   \hat{b} = \tilde b + i(\tilde a_1- \tilde a_2) \, , 
\end{align}
where we recall that $\tilde{a}_i$ was defined in \eqref{tildeai}. 
The result for $I_1$ will not have such a simple form, but it can be expressed in terms of the same two sectors. 

Finally, the  complete waveform, including contributions from both factorisation channels, is 
\begin{align}
\begin{split}
\label{G3-c}
    h^{\infty}_{G_3} (u) = h^{\infty, (1)}_{G_3} (u) + h^{\infty, (2)}_{G_3} (u)\, , 
\end{split}
\end{align}
where
\begin{align}
\label{swap}
   h^{\infty, (2)}_{G_3} (u) = \left.h^{\infty, (1)}_{G_3} (u)\right|_{(m_1, a_1,  v_1, b_1){\leftrightarrow} (m_2, a_2,  v_2, b_2)}\, .  
\end{align}
We also observe  that under this exchange the variable $\tilde{b}$ defined in \eqref{btildeetal} changes sign, $\tilde{b} \leftrightarrow  - \tilde{b}$.

\subsection{The $I_1$ waveform}\label{ssec:I1-wf}

In the presence of the $I_1$ deformation, the time-domain waveform  is 
\begin{align}
\begin{split}
\label{interI1-ter}
    &h^{\infty, (1)}_{I_1} (u) = - \frac{i \kappa}{2 m_1 m_2 (\hat{k}\Cdot v_2)}
    \int\!\frac{d^4 q_1}{(2\pi)^3}\delta(2 q_1 \Cdot v_1) e^{i q_1\Cdot \tilde{b}} \left.\cM_{I_1, q_1^2}\right|_{k = \omega^\ast \hat{k}}\, , 
    \end{split}
\end{align}
where we have, from \eqref{firstI1iint},
\begin{align}
\begin{split}
\label{eq:I1}
\left.\cM_{I_1, q_1^2}\right|_{k = \omega^\ast \hat{k}} &= \  -  i \left(\frac{\kappa}{2}\right)^5\frac{12\,   m_1^2 m_2^2\, (\omega^\ast)^3}{q_1^2\, (q_1\Cdot \hat{k})}  \, 
  (X \Cdot q_1)^2(v_2 \Cdot \hat{k} )\, e^{- a_1 \Cdot q_1} \times
  \\ & \bigg[   \cosh [ a_2\Cdot (\omega^\ast \hat{k}-q_1)]  (v_2 \Cdot \hat{k}) -  i\,    \frac{\sinh a_2\Cdot (\omega^\ast \hat{k}-q_1)}{a_2\Cdot (\omega^\ast \hat{k}-q_1)}\epsilon ( \hat{k} v_2 q_1 a_2)\bigg] \ , 
  \end{split}
\end{align}
with $\omega^\ast$  given in \eqref{omast}. 
Therefore we obtain  
\begin{align}
\begin{split}
\label{interI1-bis}
    &h^{\infty, (1)}_{I_1} (u)
    = - \left(\frac{\kappa}{2}\right)^6 
\frac{6m_1 m_2}{(\hat{k}\Cdot v_2)^3}\int\!\frac{d^4q_1}{(2\pi)^3} \delta(2 q_1\Cdot v_1) \frac{(X \Cdot q_1)^2  (q_1\Cdot v_2)^3}{q_1^2\, (q_1\Cdot \hat{k})} \times\\
    & \left[ \left( v_2 \Cdot \hat{k} + i \frac{\epsilon(\hat{k} v_2 q_1 a_2)}{q_1 \Cdot \tilde{a}_2}\right)  e^{i q_1 \Cdot [  \tilde{b} +  i (\tilde{a}_1 + \tilde{a}_2) ]} + \left( v_2 \Cdot \hat{k}  - i \frac{\epsilon(\hat{k} v_2 q_1 a_2)}{q_1 \Cdot \tilde{a}_2}\right)  e^{i q_1 \Cdot [  \tilde{b} +  i (\tilde{a}_1 -  \tilde{a}_2) ]}\right]
    \, , 
\end{split}
\end{align}
where the shifted spin vectors  $\tilde{a}_i$ were defined in \eqref{tildeai}. 

We now  derive using the  method of generating functions presented in Appendix~\ref{sec:AppB}
the two master integrals  needed to evaluate the waveform integral  in  \eqref{interI1-bis}, corresponding to the first factorisation diagram. 
Before doing so, we  comment that 
its integrand   contains the spurious pole $1/ (q_1\Cdot \hat{k})$, which originates from the propagator $i/(q^2 + i \varepsilon) =  i / ( -2 q_1\Cdot \hat{k} + i \varepsilon)$ in the Compton amplitude \eqref{I1amplitude-bis} onto which the five-point amplitude factorises. We have checked explicitly that this pole is cancelled by a corresponding spurious pole in the second factorisation diagram, and have double-checked this conclusion by a detailed comparison with the full five-point amplitude obtained from a diagrammatic computation. In practice it is convenient to perform the integrations of the two factorisation diagrams separately, which requires a choice of regularisation of the pole. The sum of the two diagrams should be independent of the  regularisation, since the pole is spurious; 
we have confirmed this expectation by performing the integration with either the Feynman $i \varepsilon$ prescription or the Principal Value (PV)  prescription,  finding complete agreement.%
\footnote{In fact, using the PV prescription the spurious contributions vanish separately in each diagram; this is related to the tracelessness of the integral $\mathcal{K}^{\mu\nu}$
introduced in \eqref{Kmunu-defint}, see end of Appendix~\ref{sec:AppA} for a thorough discussion of this point.
}
In the following we will choose the PV prescription, and for completeness    we  give in Appendix~\ref{sec:AppA} the master integral $\mathcal{J}^\mu$ in \eqref{Jmu-def} with both prescriptions (while for $\mathcal{K}^{\mu\nu}$ we only present the result in the PV prescription).

The first type of integral we need is
\begin{align}\label{I1gen-cosh}
\begin{split}
\cD^{(1)}&\coloneq\int\!\frac{d^4q_1}{(2\pi)^3} \delta(2 q_1\Cdot v_1) \frac{(q_1 \Cdot X)^2 (q_1\Cdot v_2)^3}{q_1^2 (q_1 \Cdot \hat{k})} e^{i q_1 \Cdot \tilde{b}} \ , \\
&= \left[  \frac{\partial^4}{\partial^2 t_1 \partial^2 t_2} \left(
 \mathcal{J} \Cdot v_2 \right)_{\tilde{b} \to \tilde{b}+t_1 v_{2} + t_2 X}
 \right]_{t_1=t_2=0}  \\
& = 
\left[  \frac{\partial^4}{\partial^2 t_1 \partial^2 t_2} \left(\frac{v_2 \Cdot K_1 \Cdot \hat{k}}{8 \pi |\tilde{b}_{(1)}| \, \big[(v_1 \Cdot \hat{k})^2 |\tilde{b}_{(1)}|^2-(\hat{k} \Cdot \tilde{b}_{(1)})^2 \big]} \right)_{\tilde{b} \to \tilde{b}+t_1 v_{2} + t_2 X}
 \right]_{t_1=t_2=0} \, , 
\end{split}
\end{align}
where we have used \eqref{Jmu} and 
$K_1^{\mu\nu} = |\tilde{b}_{(1)}|^2 P_1^{\mu\nu} + \tilde{b}_{(1)}^\mu 
\tilde{b}_{(1)}^\nu$ from \eqref{K1def}.

The second type is
\begin{align}\label{I1gen-cosh-bis}
\begin{split}
\mathcal{E}^{(1)}&\coloneq\int\!\frac{d^4q_1}{(2\pi)^3} \delta(2 q_1\Cdot v_1) \frac{(q_1 \Cdot X)^2 (q_1\Cdot v_2)^3 \epsilon(\hat{k} v_2 q_1 a_2)}{q_1^2 (q_1 \Cdot \hat{k})(q_1 \Cdot \tilde{a}_2)} 
e^{i q_1 \Cdot \tilde{b}}  \\
&= -\left[  \frac{\partial^4}{\partial^2 t_1 \partial^2 t_2} \left(Y \Cdot
 \mathcal{K} \Cdot v_2 \right)_{\tilde{b} \to \tilde{b}+t_1 v_{2} + t_2 X}
 \right]_{t_1=t_2=0} \ , 
\end{split}
\end{align}
with 
\begin{align}
Y \Cdot \mathcal{K} \Cdot v_2 = 
  - \frac{\big( \tilde{a}_2 \Cdot K_{1} \Cdot \hat{k} \big) (Y \Cdot K_{1} \Cdot v_2) - (\tilde{a}_2 \Cdot K_{1} \Cdot Y) (\hat{k} \Cdot K_{1} \Cdot v_2)-
(\tilde{a}_2 \Cdot K_{1} \Cdot v_2) (\hat{k} \Cdot K_{1} \Cdot Y)}{8 \pi  |\tilde{b}_{(1)}| \left( \tilde{a}_2 \Cdot K_{1} \Cdot \tilde{a}_2 \right) \left( \hat{k} \Cdot K_{1} \Cdot \hat{k} \right)} \ ,
\end{align}
where we have used
$Y^{\mu}\coloneq\epsilon(\hat k v_2\mu a_2)$, as well as \eqref{Kmunu-defint} and \eqref{Kmunu}.

Putting everything together, we find that \eqref{interI1-bis} evaluates to
\begin{align}
\begin{split}
\label{interI1-final}
    &h^{\infty, (1)}_{I_1} (u)
    = - \left(\frac{\kappa}{2}\right)^6 
\frac{6m_1 m_2}{(\hat{k}\Cdot v_2)^3} \times\\
    & \left[ \left( (\hat{k} \Cdot v_2) \cD^{(1)}+ i \mathcal{E}^{(1)}\right)_{\tilde{b} \to  \tilde{b} +  i (\tilde{a}_1 + \tilde{a}_2)} + 
\left( (\hat{k} \Cdot v_2) \cD^{(1)}- i \mathcal{E}^{(1)}\right)_{\tilde{b} \to  \tilde{b} +  i (\tilde{a}_1 - \tilde{a}_2)}    
    \right]
    \, . 
\end{split}
\end{align}
As for the $G_3$ case, the complete $I_1$ waveform is 
\begin{align}
\begin{split}
\label{I1-c}
    h^{\infty}_{I_1} (u) = h^{\infty, (1)}_{I_1} (u) + h^{\infty, (2)}_{I_1} (u)\, , 
\end{split}
\end{align}
where
\begin{align}
\label{swap-2}
   h^{\infty, (2)}_{I_1} (u) = \left.h^{\infty, (1)}_{I_1} (u)\right|_{(m_1, a_1,  v_1, b_1){\leftrightarrow} (m_2, a_2,  v_2, b_2)}\, .  
\end{align}
Finally, we  note the appearance in \eqref{interI1-bis} of shifted impact parameters $\tilde{b} +  i (\tilde{a}_1 \pm  \tilde{a}_2) $,  similarly to \eqref{interG3} in the $G_3$ case. However there is an important difference with that case: while  the $G_3$ waveform can be obtained from the scalar {\it integrated} waveform by performing shifts in the impact parameter as dictated by the sectors, in the $I_1$ case these shifts occur at the level of the {\it integrand}. For the $\cosh$ part of the $I_1$ integrand they translate to shifts that can be performed on the integrated result; however the part of the $I_1$ proportional to the $\sinh$ function vanishes in the scalar case, and thus provides a new contribution that cannot be obtained from the spinless waveform.

\subsection{Waveforms for parity-odd deformations}

To derive the waveforms for parity-odd deformations we only need to recall from Section~\ref{sec:Compton} that the Compton amplitudes for parity-odd deformations are obtained from the parity-even ones by multiplying them by a factor of $\pm i$ depending on whether the two gravitons have positive or negative helicity. The same property is inherited by the cut five-point amplitudes which we use to construct the waveforms, that is 
\begin{align}
    -i \cM_5^{\rm P.O.} (k^+) = i \big(-i \cM_5^{\rm P.E.}(k^+)\big)\, , \qquad 
     \big(-i \cM_5^{\rm P.O.} (k^-)\big)^\ast = i \big(-i \cM_5^{\rm P.E.}(k^-)\big)^\ast\, , 
\end{align}
from which it follows that 
\begin{align}
\label{prima}
    W^{\rm P.O.} (k^+) = i W^{\rm P.E.}(k^+)\, , \qquad 
     \big(W^{\rm P.O.} (k^-)\big)^\ast = i \big(W^{\rm P.E.}(k^-)\big)^\ast\, , 
\end{align}
where P.O.~and P.E.~stand for parity odd and even, respectively.
Combining \eqref{prima} with \eqref{seconda} 
we see that 
\begin{align}
\left.    \big(W^{\rm P.O.} (k^-)\big)^\ast\right|_{k = - \omega (1,\mathbf{\hat{x}}) } = \left.W^{\rm P.O.} (k^+)\right|_{k =  \omega (1,\mathbf{\hat{x}}) } = 
\left.i  W^{\rm P.E.} (k^+)\right|_{k = \omega (1,\mathbf{\hat{x}}) }\, . 
\end{align}
The two terms in  
\eqref{KMOCsubfinalforhmunuquater}, can then be combined as in the parity-even case, 
and hence we conclude that 
\begin{align}
\begin{split}
    (h^{\infty, \rm P.O.}_{+} + i\,   h^{\infty, \rm P.O.}_{\times}) &= 
    \kappa \int_{-\infty}^{+\infty}\!
\frac{d\omega}{2\pi}e^{- i \omega u} 
 \left. 
W^{\rm P.O.}(b, k)\right|_{k=\omega(1, \hat{\mathbf{x}})}
\\ 
&=
    i \, (h^{\infty, \rm {P.E.}}_{+} + i\,   h^{\infty, \rm P.E.}_{\times})\, . 
    \end{split}
\end{align}
This implies that  
\begin{align}
\label{pcswap}
    h^{\infty, \rm P.O.}_{+} = -    h^{\infty, \rm P.E.}_{\times}\, , \qquad  
    h^{\infty, \rm P.O.}_{\times} = h^{\infty, \rm P.E.}_{+}\, , 
\end{align}
that is,  the ``plus'' and ``cross'' polarisations are then swapped  in the way prescribed by \eqref{pcswap}.

\section{Time-domain waveforms from  tensor integral generating functions, reloaded}
\label{sec:TIGF}
Alternatively, in the spirit of modern multiloop amplitude calculations, we can compute the integrals by a systematic method proposed in \cite{Chen:2024bpf}.
For the tree-level waveform, we have two $D$-dimensional master integrals,    
\begin{align}
\begin{split}\label{eq:tigfMaster}
   \cI_1[\mathbf{y}]\coloneq \int\!\frac{d^Dq_1}{(2\pi)^{D-1}}  {\delta(2 q_1\Cdot v_1)\over q_1^2}  e^{i q_1 \Cdot \hat{b}}\,, ~~~~~
\cI_2[\mathbf{y}]\coloneq\int\!\frac{d^{D}q_1}{(2\pi)^{D-1}}  {\delta(2 q_1\Cdot v_1)\over q_1^2 q_1\mdot \hat k}  e^{i q_1 \Cdot \hat{b}} \,,
    \end{split}
\end{align}
where $\mathbf{y}\coloneq (y_1, y_2, y_3,\hat y_4,\hat y_5,\hat y_6,\hat y_7)$, with 
\begin{align}
\label{yss}
    y_1&\coloneq\sigma=v_1\mdot v_2, & y_2&\coloneq v_1\mdot \hat k , &y_3&\coloneq v_2\mdot \hat k , \nn\\
     \hat y_4&\coloneq {\hat b\mdot \hat k \over \sqrt{-\hat b \mdot \hat b}}=0, & \hat y_5&\coloneq {\hat b\mdot v_1 \over \sqrt{-\hat b \mdot \hat b}}, &\hat y_6&\coloneq {\hat b\mdot v_2\over \sqrt{-\hat b \mdot \hat b}}, & \hat y_7\coloneq \hat b \mdot \hat b\,. 
\end{align}
One can rewrite the two master integrals as 
\begin{align}
	\cI_{j}[\mathbf{y}]\coloneq\int_{-\infty}^{\infty}\!dt \, e^{it} \ \widetilde{\cI}_{j}[\mathbf{y}, t]&:=\int_{-\infty}^{\infty}\!dt\,  e^{it} \int\!\frac{d^{D-4}q_1}{(2\pi)^3}  {\delta(2 q_1\Cdot v_1)\over q_1^2 (q_1\mdot k)^{j-1}}  \delta(q_1\mdot \hat b-t)
 \, . 
 %\nn\\
	%&=\Big(\int_{-\infty}^{\infty}\!dt \, e^{it} \ {t^{-3-j + D} \over D-4} \Big)\widehat{\cI}_{j}[\mathbf{y}]\, .
\end{align}
The differential equation in $t$ for $\widetilde{\cI}_{j}[\mathbf{y}, t]$ is very simple:
\begin{align}
    \partial_t  \widetilde{\cI}_{1}[\mathbf{y}, t]= \frac{D-4}{t} \widetilde{\cI}_{1}[\mathbf{y}, t], && \partial_t  \widetilde{\cI}_{2}[\mathbf{y}, t]= \frac{D-5}{t} \widetilde{\cI}_{2}[\mathbf{y}, t]\,.
\end{align}
It is easy to see that  the $t$ dependence is factorised from the other variables, and we can solve for it directly as 
\begin{align}
    \widetilde{\cI}_{1}[\mathbf{y}, t]= {t^{D-4}\over D-4} \widehat{\cI}_{1}[\mathbf{y}]\,, && \widetilde{\cI}_{2}[\mathbf{y}, t]= {t^{D-5}} \widehat{\cI}_{2}[\mathbf{y}]\, . 
\end{align}
Then the two integrals in \eqref{eq:tigfMaster} are rewritten as  
\begin{align}
	\cI_{1}[\mathbf{y}]&=\Big(\int_{-\infty}^{\infty}\!dt \, e^{it} \, {t^{-4 + D} \over D-4} \Big)\widehat{\cI}_{1}[\mathbf{y}]=-\pi \, \widehat{\cI}_{1}[\mathbf{y}]\, ,\nn\\ 
  \cI_{2}[\mathbf{y}]&=\Big(\int_{-\infty}^{\infty}\!dt \, e^{it} \, {t^{-5 + D} } \Big)\widehat{\cI}_{2}[\mathbf{y}]=i\pi \, \widehat{\cI}_{2}[\mathbf{y}]\, , 
\end{align}
where we have already used $D = 4 - 2\epsilon$ and taken the $\eps{\to} 0$ limit.
The differential equations in four  spacetime dimensions  of the  $\mathbf{y}$-dependent part $\widehat{\cI}_{1,2}[\mathbf{y}]$ are 
\begin{align}
	\partial_{\hat y_5}\widehat{\cI}_{1}[\mathbf{y}]= -\frac{\hat y_5}{\hat y_5^2+1} \widehat{\cI}_{1}[\mathbf{y}], &&
	\partial_{y_2}\widehat{\cI}_{2}[\mathbf{y}]= -{1\over  y_2}\widehat{\cI}_{2}[\mathbf{y}].
\end{align}
The solutions  are 
\begin{align}
	\widehat{\cI}_{1}[\mathbf{y}]=\frac{-\pi}{(2\pi)^3\sqrt{-\hat b\mdot \hat b}\sqrt{\hat y_5^2+1}},&&
	\widehat{\cI}_{2}[\mathbf{y}]={c_2\over  y_2}\,, 
\end{align}
where $c_2=0$ from the boundary conditions. 
Hence we can omit $\widehat{\cI}_2[\mathbf{y}]$ in the calculation and we are thus left with a single master integral. 

All other integrals can be reduced to this master integral. For example,  consider the case of the integral
\begin{align}
     \widehat{\mathcal{V}}[\mathbf{y}]&\coloneq  \int\!\frac{d^{D}q_1}{(2\pi)^{D-1}}  {\delta(2 q_1\Cdot v_1)q_1\mdot v_2\over q_1^2 q_1\mdot \hat k}  e^{i q_1 \Cdot \hat{b}} \, .
\end{align}
We can rewrite it as 
\begin{align}
	\widehat{\mathcal{V}}[\mathbf{y}]=\int_{-\infty}^{\infty}dt \ e^{it} \ \widetilde{\mathcal{V}}[\mathbf{y}, t]&=\int_{-\infty}^{\infty}\!dt \, e^{it}\int\!\frac{d^{D}q_1}{(2\pi)^{D-1}}  {\delta(2 q_1\Cdot v_1)q_1\mdot v_2\over q_1^2 (q_1\mdot k)}  \delta(q_1\mdot \hat b-t)\, . 
 %\nn\\
%	&=\int_{-\infty}^{\infty}\!dt\,  e^{it}  J_{1,1,1,1,-1}\, .
\end{align}
By  IBP reduction, the $q_1$-integral is reduced to the single master integral  $\widehat{\cI}_{1}[\mathbf{y}]$, and finally we get 
%\begin{align}
%	J_{1,1,1,1,-1}=\frac{\left(y_1 \left(y_2+\hat y_4 \hat y_5\right)-y_3 \left(\hat y_5^2+1\right)-\hat y_4 \hat y_6+y_2 \hat y_5 \hat y_6\right) }{y_2^2+2 \hat y_4 \hat y_5 y_2-\hat y_4^2} J_{1,0,1,1,0}.
%\end{align}
%Finally we get
\begin{align}
	\widehat{\mathcal{V}}[\mathbf{y}]
	&=(-\pi)\frac{y_1 \left(y_2+\hat y_4 \hat y_5\right)-y_3 \left(\hat y_5^2+1\right)-\hat y_4 \hat y_6+y_2 \hat y_5 \hat y_6}{y_2^2+2 \hat y_4 \hat y_5 y_2-\hat y_4^2}\widehat{\cI}_{1}[\mathbf{y}]\, . 
\end{align}
We have checked that this formula is  consistent with \eqref{Jmu}.   
%The $\cI_2[\mathbf{y}]$ can not be taken as the TIGF due to the trivial dependent on $\hat b$. 
%Then 
%\begin{align}
%	\frac{c_1}{\sqrt{-\hat b\mdot \hat b}\sqrt{y_5^2+1}}\frac{1}{y_2^2+2 y_4 y_5 y_2-y_4^2}
%\Big(y_1 \left(y_2+y_4 y_5\right)-y_3 \left(y_5^2+1\right)-y_4 y_6+y_2 y_5 y_6 \Big)
%\end{align}

 The integral in \eqref{eq:I1} contains a spurious pole in the entire function $(\sinh x)/x$ (see \cite{Bjerrum-Bohr:2023iey,Bjerrum-Bohr:2023jau} for general discussions). However we can eliminate it  by introducing the integral representation   \cite{Chen:2024bpf} 
\begin{align}
\label{trick}
    {\sinh x \over x}= \int_0^1\!dz \, \cosh(z\, x)\, . 
\end{align}
Then there are four sectors in the waveform, with exponential factors $e^{i q \mdot \hat{b} }$. The particular  form of the new variable $\hat b$ depends on the  sector as follows:%
\footnote{We note that in the previous approach we employed to compute the $I_1$ waveform there were  two sectors, corresponding to the cases  {\bf (1)} and {\bf (2)} below. We now have two additional sectors because 
 the integral representation    \eqref{trick} introduces a   dependence on the integration variable~$z$.}
\begin{align}
    {\bf (1)}~~~~~~~&& {\bf (2)} ~~~~~~~&& {\bf (3)} ~~~~~~~&& {\bf (4)} ~~~~~~~\nn\\
   \hat b= \tilde b+i(\tilde a_1+ \tilde a_2)  &&   \hat b=\tilde b+i(\tilde a_1- \tilde a_2) && \hat b= \tilde b+i(\tilde a_1+ z \tilde a_2) && \hat b= \tilde b+i (\tilde a_1- z \tilde a_2)\, .
\end{align}
Then the waveform can be represented as a combination of tensor integrals:
    \begin{align}
\begin{split}
\label{eq:wffinal}
    &h^{\infty, (1)}_{G_3} (u)
    = - \left(\frac{\kappa}{2}\right)^6
\frac{6m_1 m_2}{(\hat{k}\Cdot v_2)^3} \Big({1\over 4}X^{\mu_1}X^{\mu_2}v_2^{\mu_3}v_2^{\mu_4} \sum_{i=1}^2 \cI^{(i)}_{\mu_1\mu_2\mu_3 \mu_4}(y)\Big) \, ,\\
&h^{\infty, (1)}_{I_1} (u)
    = - \left(\frac{\kappa}{2}\right)^6
\frac{6m_1 m_2}{(\hat{k}\Cdot v_2)^3} \Big(v_2\mdot \hat k X^{\mu_1}X^{\mu_2}v_2^{\mu_3}v_2^{\mu_4} \sum_{i=1}^2 \cI'^{(i)}_{\mu_1\mu_2\mu_3 \mu_4}(y)\\
&~~~~~~~~~~~~~~~~~~~~~~~~~~~~~~~~~~~+X^{\mu_1}X^{\mu_2}v_2^{\mu_3}v_2^{\mu_4}Y^{\mu_5} \sum_{i=3}^4 \cI'^{(i)}_{\mu_1\mu_2\mu_3 \mu_4\mu_5}(y)\Big) 
\, , 
\end{split}
\end{align}
where $Y^{\mu}\coloneq\epsilon(\hat k v_2\mu a_2)$,  and  $y\coloneq (y_1, y_2, y_3, y_4, \ldots, y_{18})$  denotes all the independent scalar products among external kinematic vectors $v_1, v_2, \hat k, \tilde a_1, \tilde a_2, \tilde b$, with $y_1, y_2, y_3$ defined in \eqref{yss} and 
\begin{align}
      y_4&\coloneq {\tilde b\mdot \hat k \over \sqrt{-\tilde b \mdot \tilde b}}=0, &  y_5&\coloneq {\tilde b\mdot v_1 \over \sqrt{-\tilde b \mdot \tilde b}}, & y_6&\coloneq {\tilde b\mdot v_2\over \sqrt{-\tilde b \mdot \tilde b}}, &  y_7&\coloneq \tilde b \mdot \tilde b\,,  \nn\\
      y_8&\coloneq {\tilde a_1\mdot \hat k \over \sqrt{-\tilde b \mdot \tilde b}}=0, & y_9&\coloneq {\tilde a_1\mdot v_1\over \sqrt{-\tilde b \mdot \tilde b}}, & y_{10}&\coloneq {\tilde a_1\mdot v_2\over \sqrt{-\tilde b \mdot \tilde b}}, &  y_{11}&\coloneq {\tilde a_1 \mdot \tilde a_1\over -\tilde b \mdot \tilde b}\, & \nn \\ 
       y_{12}&\coloneq {\tilde a_2\mdot \hat k \over \sqrt{-\tilde b \mdot \tilde b}}=0,  & y_{13}&\coloneq {\tilde a_2\mdot v_1\over \sqrt{-\tilde b \mdot \tilde b}}, &  y_{14}&\coloneq {\tilde a_2\mdot v_2\over \sqrt{-\tilde b \mdot \tilde b}}, & y_{15}&\coloneq {\tilde a_2 \mdot \tilde a_2\over -\tilde b \mdot \tilde b}\, & \nn\\ 
       y_{16}&\coloneq {\tilde a_1 \mdot \tilde b\over -\tilde b \mdot \tilde b},& y_{17}&\coloneq {\tilde a_2 \mdot \tilde b\over -\tilde b \mdot \tilde b},& y_{18}&\coloneq {\tilde a_1 \mdot \tilde a_2\over -\tilde b \mdot \tilde b}\, . 
\end{align}
The tensor integrals are generated as  
\begin{align}
\begin{split}
\label{eq:tigf0}
 \cI^{(i)}_{\mu_1\ldots \mu_r}(y)&= \partial_{\tilde b^{\mu_1}}\cdots \partial_{\tilde b^{\mu_r}}\ \cI_{1}^{(i)}[\mathbf{y}]= \partial_{\tilde b^{\mu_1}}\cdots \partial_{\tilde b^{\mu_r}}\int_{0}^1dz \ \widehat{\cI_{1}}^{(i)}[\mathbf{y}]\, ,  \\
 \cI'^{(i)}_{\mu_1\ldots \mu_r}(y)&= \partial_{\tilde b^{\mu_1}}\cdots \partial_{\tilde b^{\mu_r}}\ \mathcal{V}^{(i)}[\mathbf{y}]=\partial_{\tilde b^{\mu_1}}\cdots \partial_{\tilde b^{\mu_r}}\int_{0}^1dz\ \widehat{\mathcal{V}}^{(i)}[\mathbf{y}]\, . 
\end{split}
\end{align}
We note that one cannot use $\widehat{\cI}_{2}$ to generate the tensor integral as its value is trivial. Furthermore,  it is not possible to   set $y_4{=}0$ in $\widehat{\mathcal{V}}^{(i)}$ and $\widehat{\cI}^{(i)}_{1}$ directly,  as $\partial_{\tilde{b}^\mu}y_4$ is nonvanishing.  The superscripts in $\widehat{\mathcal{V}}^{(i)}$ and $\widehat{\cI}^{(i)}_{1}$ denote the tensor generating function with $\hat b$ in different sectors. Then we only need to perform the $z$ integration for the simple cases of $\widehat{\cI_{1}}^{(3,4)}[\mathbf{y}]$ and $\widehat{\mathcal{V}}^{(3,4)}[\mathbf{y}]$. By using IBP and partial fractioning, the $z$ integral can  then  be reduced  to a basis of  three integrals:
\begin{align}
\label{eq:master2-bis}
\int_0^1\!  \, \frac{dz}{ \mathcal{Y}}, &&\int_0^1\!  \, \frac{dz}{\hat b\mdot \hat b \, \mathcal{Y}}, &&\int_0^1\!  \, \frac{ dz\, z}{ \hat b\mdot \hat b\, \mathcal{Y}}\, , 
\end{align}
where  $\mathcal{Y}^2\coloneq  -\hat b\mdot \hat b \, \hat y_5^2-\hat b\mdot \hat b$, which 
 are straightforward to evaluate directly.

%The evaluation of $\sigma$ variable are easy to perform directly. 

%\begin{align}
%\begin{split}
%\label{interG3-bis}
%    &h^{\infty, (1)}_{G_3} (u) = - \frac{i \kappa}{2 m_1 m_2 (\hat{k}\Cdot v_2)}
%    \int\!\frac{d^4 q_1}{(2\pi)^3}\delta(2 q_1 \Cdot v_1) e^{i q_1\Cdot \tilde{b}} \left.\cF_{G_3}^{(1)}\right|_{k = \omega^\ast \hat{k}}
%    \\ &=- 6 m_1 m_2 \left(\frac{\kappa}{2}\right)^6
%    \frac{[\hat{k}| v_1 \sigma_\mu |\hat{k}] [\hat{k}| v_1 \sigma_\nu |\hat{k}]
%    }{(\hat{k}\Cdot v_2)^2}\cdot \frac{1}{2 (\hat{k}\Cdot v_2)}\int\!\frac{d^4q_1}{(2\pi)^3} \delta(2 q_1\Cdot v_1) \frac{q_1^\mu q_1^\nu (q_1\Cdot v_2)^2}{q_1^2} \times\\
%    & \ \ \ \ \frac{1}{2}\left[ e^{i q_1 \Cdot [  \tilde{b} +  i (\tilde{a}_1 - \tilde{a}_2) ]} + e^{i q_1 \Cdot [  \tilde{b} + i (\tilde{a}_1 + \tilde{a}_2)]}\right]. 
    %& = - \frac{3}{2} m_1 m_2 \left(\frac{\kappa}{2}\right)^6
    %\frac{[\hat{k}| v_1 \sigma_\mu |\hat{k}] [\hat{k}| v_1 \sigma_\nu |\hat{k}] v_{2 \alpha} v_{2\beta}
    %}{(\hat{k}\Cdot v_2)^3}\Big[ 
    %\left.\cI^{\mu \nu \alpha \beta}\right|_{\tilde{b}\to \tilde{b} + i(\tilde{a}_1 - \tilde{a}_2)} + \left.\cI^{\mu \nu \alpha \beta}\right|_{\tilde{b}\to \tilde{b}+ i(\tilde{a}_1 + \tilde{a}_2)}\Big] 
    %\, , 
%\end{split}
%\end{align}
%DONE!
\section{Time-domain waveforms from residues }
\label{sec:residues}
A final approach, followed in \cite{DeAngelis:2023lvf,Brandhuber:2023hhl}, 
makes use of Cauchy's residue theorem. We start by rewriting the original expression for the waveform in \eqref{KMOCsubfinalbis2} as follows,
\begin{align}
\begin{split}
 h_I^\infty(u) 
&{=} -i\kappa \int_{-\infty}^{+\infty}\!
\frac{d\omega}{2\pi}e^{- i \omega (u-\hat{k}\Cdot\bt_2)} 
   \int\!  \frac{d^4q_1}{(2\pi)^{2}}  \delta(2  {p}_1\Cdot q_1 ) \delta(2  {p}_2\Cdot (k-q_1) )\  e^{iq_1\Cdot \bt} \ \mathcal{M}_I \, ,
\end{split}
\end{align}
where $k=\omega \hat{k}$. Further, we  split up the waveform into contributions coming from the two cuts \eqref{2factordiagrams2}, and rescale the momentum transfers as $q_i=\omega \hat{q}_i$,  
\begin{align}
\begin{split}
h_I^{\infty, (i)} (u) 
{=} -i\kappa \int\!  \frac{d^4\hat{q}_1}{(2\pi)^{2}}\int_{-\infty}^{+\infty}\!
\frac{d\omega}{2\pi}\, &\omega^2 e^{- i \omega (u-\hat{k}\Cdot\bt_2 -\hat{q}_1\Cdot \bt)} 
    \delta(2  {p}_1\Cdot \hat{q}_1 ) \delta(2  {p}_2\Cdot (\hat{k}-\hat{q}_1) )\, \\
    &\mathcal{M}_{I,q_i^2}(\omega\hat{k},\omega\hat{q}_1,\omega\hat{q}_2) .
\end{split}
\end{align}
The basic idea now is to rewrite the integral over $q_1$ as a contour integral encircling the physical poles. To do this, it is worth introducing a parameterisation which is best suited to the various sectors of the problem. 
Explicitly, we can factor out the exponential dependence on the spin for each cut, in a similar manner to \eqref{interI1-bis},
\begin{equation}\label{eq: residuceCutDecomp}
    \cM_{I,q_1^2}=e^{-\at_1 \Cdot q_1}\sum_{l=\pm}e^{l\, \at_2\Cdot q_2}\cM_{I,q_1}^{(l)}\,,\qquad  \cM_{I,q_2^2}=e^{-\at_2 \Cdot q_2}\sum_{l=\pm}e^{l\, \at_1\Cdot q_1}\cM_{I,q_2}^{(l)}\,.
\end{equation}
Focusing on the $q_1$ cut, this allows us to rewrite the waveform as 
\begin{align}
\begin{split}
\label{intab}
h_I^{\infty, (1)} (u) 
&{=} -i\kappa \int\!  \frac{d^4\hat{q}_1}{(2\pi)^{2}}\int_{-\infty}^{+\infty}\!
\frac{d\omega}{2\pi}\omega^2  
    \delta(2  {p}_1\Cdot \hat{q}_1 ) \delta(2  {p}_2\Cdot (\hat{k}-\hat{q}_1) )\   \\
    &\quad\sum_{l=\pm}e^{- i \omega (u-\hat{k}\Cdot\bt_2 -\hat{q}_1\Cdot \bt-i\at_1\Cdot \hat{q}_1+ i l\at_2\Cdot(\hat{k}-\hat{q}_1))}\mathcal{M}_{I,q_1^2}^{(l)}(\omega\hat{k},\omega\hat{q}_1,\omega\hat{q}_2)\\
&{=} -i\kappa \int\!  \frac{d^4\hat{q}_1}{(2\pi)^{2}}\int_{-\infty}^{+\infty}\!
\frac{d\omega}{2\pi}\omega^4  
    \delta(2  {p}_1\Cdot \hat{q}_1 ) \delta(2  {p}_2\Cdot (\hat{k}-\hat{q}_1) )\   \\
    &\quad\sum_{l=\pm}e^{- i \omega (\fu_l-\fb_l\Cdot q_1)}\mathcal{M}_{I,q_1^2}^{(l)}(\hat{k},\hat{q}_1,\hat{q}_2)\, ,
\end{split}
\end{align}
where we have extracted the $\omega$ dependence of $\mathcal{M}_{I,q_i^2}^{(l)}$, which is simply a factor of $\omega^2$, and we have defined
\begin{equation}
    \fu_l\coloneqq u-\hat{k}\Cdot\bt_2+ i \, l\, \at_2\Cdot\hat{k}\, , \quad \fb_l^\mu\coloneq\bt^\mu + i \at_1^\mu +i \, l\, \at_2^\mu\,.
\end{equation}
To evaluate \eqref{intab}, we parameterise the $q_1$ integral in each sector using~\cite{Cristofoli:2021vyo}
\begin{align}
    z_1\coloneqq v_1\Cdot q_1, \quad z_2\coloneqq& v_2\Cdot q_1, \quad  z_{\fb_l}\coloneqq \fb_l\Cdot q_1, \quad z_{o_l}\coloneqq o_{l}\Cdot q_1\,,\end{align}
    with 
    \begin{align}
    &o_{l}^\mu\coloneq \epsilon(v_1 v_2 \bt_l\mu )\,.
\end{align}
The Jacobian for this transformation is simply 
\begin{equation}
    \left | \frac{\partial q_1^\mu}{\partial z_{j}}\right |= \frac{1}{|o_{l}\Cdot o_{l}|}\,.
\end{equation}
With this, the $q_1$ cut contribution to the waveform becomes
\begin{align}\label{eq: waveformResidueCut1}
\begin{split}
    h_I^{\infty, (1)} (u)&= -i\frac{\kappa}{(4\pi)^2 m_1 m_2}\left(\frac{\partial}{\partial u}\right)^4 \sum_{l=\pm}\int \!\frac{d^4 z_j}{|o_{l}\Cdot o_{l}|} \delta(z_1) \delta(z_2-v_2\Cdot\hat{k})\delta(z_{\fb_l}-\fu_l) \cM_{I,q_1^2}^{(l)}\\
    &=-i\frac{\kappa}{(4\pi)^2 m_1 m_2}\left(\frac{\partial}{\partial u}\right)^4 \sum_{l=\pm}\int_{-\infty}^{\infty} \!\frac{d z_{o_l}}{|o_{l}\Cdot o_{l}|}\cM_{I,q_1^2}^{(l)}\Big|_{z_1\rightarrow0,\,z_2\rightarrow v_2\Cdot\hat{k},\,z_{\fb_l}\rightarrow \fu_l}\,.\\
\end{split}
\end{align}
The final integral in $z_{o_l}$ can be computed using Cauchy's residue theorem. There are (at most) three poles present in $\cM_{I,q_1^2}^{(l)}$:
\begin{equation}
\begin{split}
    &\text{Physical Pole:} \quad\,\,\frac{1}{q_1^2}\sim \frac{1}{(z_{o_l}-A)(z_{o_l}-A^\star)}\,,\quad\\
    &\text{Spurious Poles:}  \quad\frac{1}{q_1\Cdot k}\sim\frac{1}{(z_{o_l}-B)}\,,\quad \frac{1}{(k-q_1)\Cdot a_2}\sim\frac{1}{(z_{o_l}-C)}\,,\\
\end{split}
\end{equation}
where $A,B,C$ are functions of the external kinematics. The residues in $1/q_1\Cdot k$ and $1/(k-q_1)\Cdot a_2$ must cancel upon summing over the sectors $l$ and cuts since the poles are spurious.\footnote{ More precisely the residue in  $1/(k-q_1)\Cdot a_2$ must cancel when summing over the sectors $l$ in \textit{each} cut. The residue in $1/q_1\Cdot k$ will only cancel after summing both cut contributions.} Thus, in practice, we only need to compute the residue on the physical pole $1/q_1^2$. This way of computing the integrals also makes it clear why terms 
with only spurious poles do not contribute to the waveform: they have vanishing residues. 

One may check that the integral \eqref{eq: waveformResidueCut1} has no pole at infinity (after applying the $\partial/\partial u$ derivatives) and that the integrand falls off sufficiently fast at infinity to close the contour above or below the axis. This contrasts with the case of  Einstein-Hilbert gravity \cite{DeAngelis:2023lvf,Brandhuber:2023hhl} where the pole at infinity exists and must be evaluated using a principal value prescription. In fact, we could have predicted this from the start. In  \cite{DeAngelis:2023lvf}, it was shown that the pole at infinity is given by the leading soft theorem of the five-point amplitude. Our five-point amplitude is built from higher-derivative corrections to general relativity and cannot change this universal leading soft theorem. Indeed, we can check that our amplitude vanishes in the soft limit $\omega{\rightarrow} 0$, which manifests in the time-domain waveform as vanishing linear memory, as discussed in Section~\ref{sec:memory}. 

To compute the contribution of the $q_2$ cut, we follow an identical procedure to the above but using the sectors for $\cM_{q_2^2}$ in \eqref{eq: residuceCutDecomp}. Upon summing both contributions, we obtain the full waveform for $I_1$ and $G_3$:
\begin{equation}
    h_I^{\infty} (u)= h_I^{\infty, (1)} (u)+h_I^{\infty, (2)} (u)\,.
\end{equation}
The expressions obtained using this method are lengthier than those found in Section~\ref{sec:directint}. Reassuringly, they are in complete agreement.

%\newpage

\section{Showcase of waveforms}
\label{sec:pictures}

In this section we show several  plots of the waveforms for the  $G_3$ and $I_1$ deformations, with and without spin. We recall that the waveforms for the parity-odd interactions $\tilde{G}_3$ and $\tilde{I}_1$ are related to the parity-even ones by \eqref{pcswap}, that is a simple swap of their real and imaginary parts; thus  separate plots for the case of parity-odd deformations are not needed. 
We will   parameterise our external kinematics as follows:%
\footnote{Our waveforms are valid for arbitrary spin configurations; however in the figures shown in this section we will focus on the case where the spins of the two bodies are aligned.}
\begin{align}
    &v_1^\mu=(\sigma,0,0,\sqrt{
\sigma^2-1})\,, &  v_2^\mu&=(1,0,0,0)\,,\nn\\
    &\hat{k}^\mu=(1,\sin \theta  \cos \phi ,\sin\theta  \sin \phi ,\cos \theta )\,,  &b^\mu&=b(0, 1, 0, 0)\,, \nn\\
   &a_1^\mu=a_1(0,0,1,0)\,, & a_2^\mu&=a_2(0,0,1,0)\,, \\
   &
   [k|_{\dot{\alpha}}=\sqrt{2}\begin{pmatrix}
\sin{\dfrac{\theta}{2}}\\[7pt] 
{-\cos{\dfrac{\theta}{2}}\, e^{-i\phi}}
\nn & &
\end{pmatrix}\, . 
%[k|_{\dot{\alpha}}=\sqrt{2}\left(\atopp{\sin{(\theta/2)}}{-\cos{(\theta/2)}e^{-i\phi}}\right)\,.  \nn & &
\end{align}
For the graphs below, we will choose $\theta{=}\phi{=}\pi/3$ and fix the spins to be either aligned or anti-aligned with the orbital angular momentum. We will also set $b {=} 1$ which means in practice 
that the spins $a_i$ and the retarded time $u$ are measured in units of $b$. Finally we will ignore a prefactor so that the waveforms are plotted in units of $\kappa^6 m_1 m_2 \beta_i$.

\subsection{Spinless case}

We begin by showing the $G_3$ and $I_1$ waveforms for the scalar case, for various values of $\sigma=v_1\Cdot v_2$.

\begin{figure}[ht]
\begin{center}
\scalebox{0.6}{\includegraphics{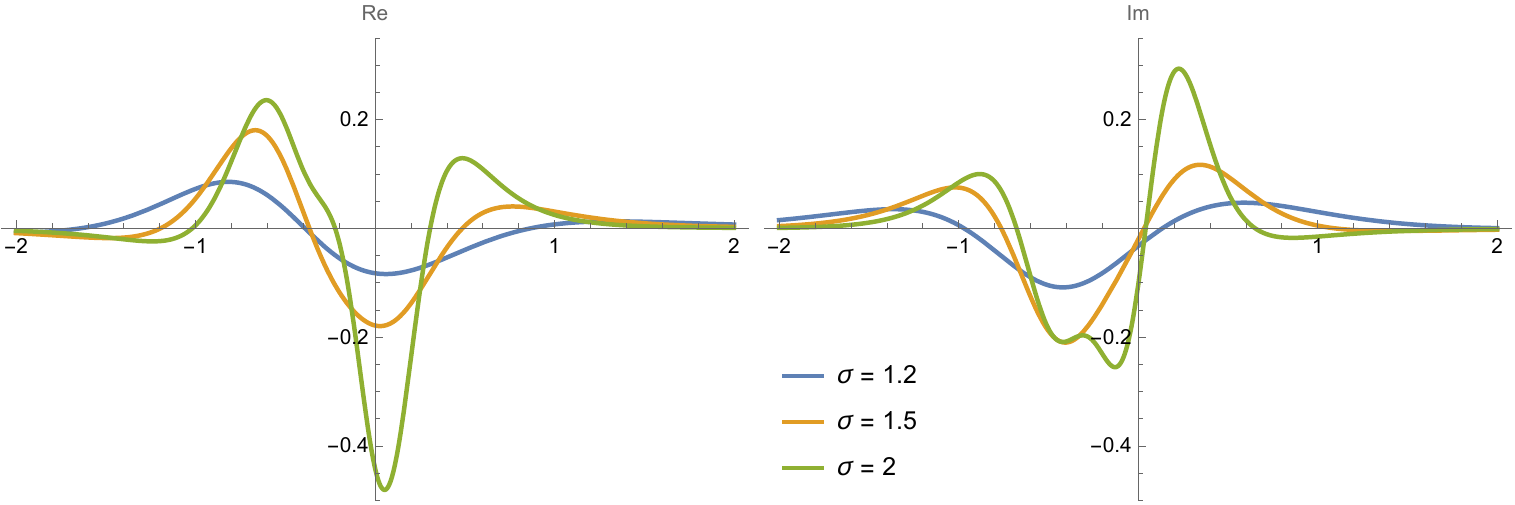}}
\end{center}
\caption{The $G_3$ waveform $(h^{\infty}_{+} + i  h^{\infty}_{\times})(u)$ plotted in the spinless case for various values of $\sigma{=}v_1\Cdot v_2$. We show separately the real and imaginary part of the waveform (the plus and cross polarisations). 
Note the absence of gravitational memory  in both the real and imaginary part of the waveform.}
\label{fig:no-spin-G_3}
\end{figure}

\begin{figure}[ht]
\begin{center}
\scalebox{0.6}{\includegraphics{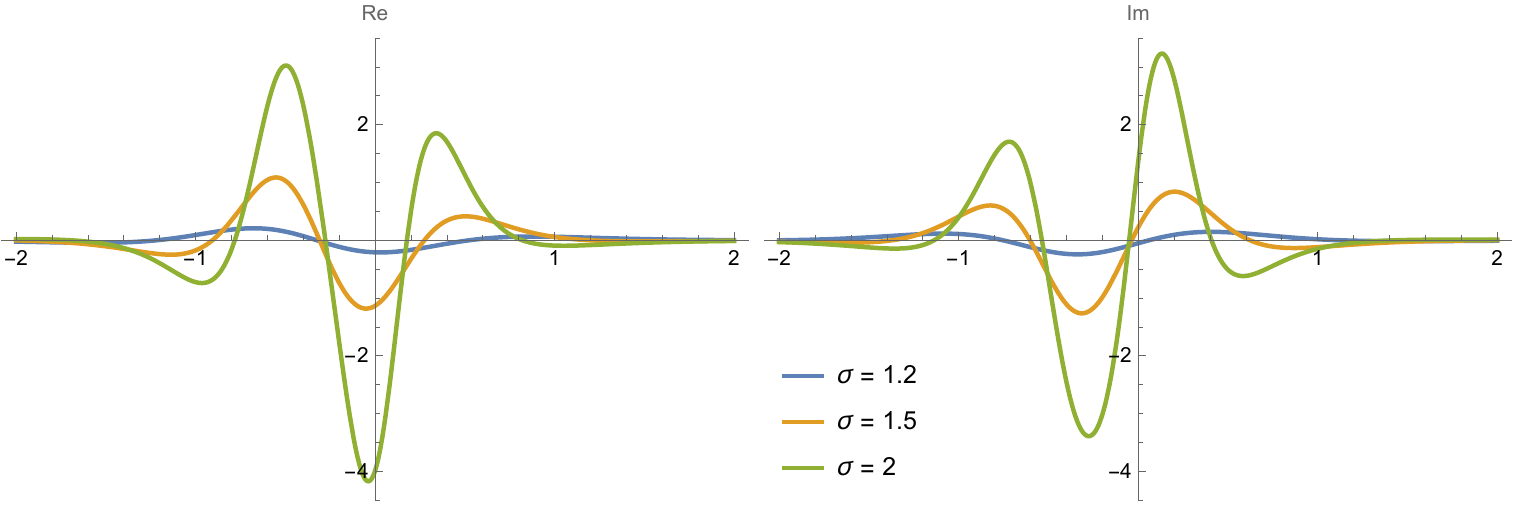}}
\end{center}
\caption{The $I_1$ waveform $(h^{\infty}_{+} + i  h^{\infty}_{\times})(u)$ plotted in the spinless case for various values of $\sigma{=}v_1\Cdot v_2$. Note that  the amplitude for  the $I_1$ deformations is about ten times larger than for the $G_3$ case.
}
\label{fig:no-spin-I_1}
\end{figure}

\FloatBarrier

\subsection{Spinning case}
\begin{figure}[ht]
\begin{center}
\scalebox{0.6}{\includegraphics{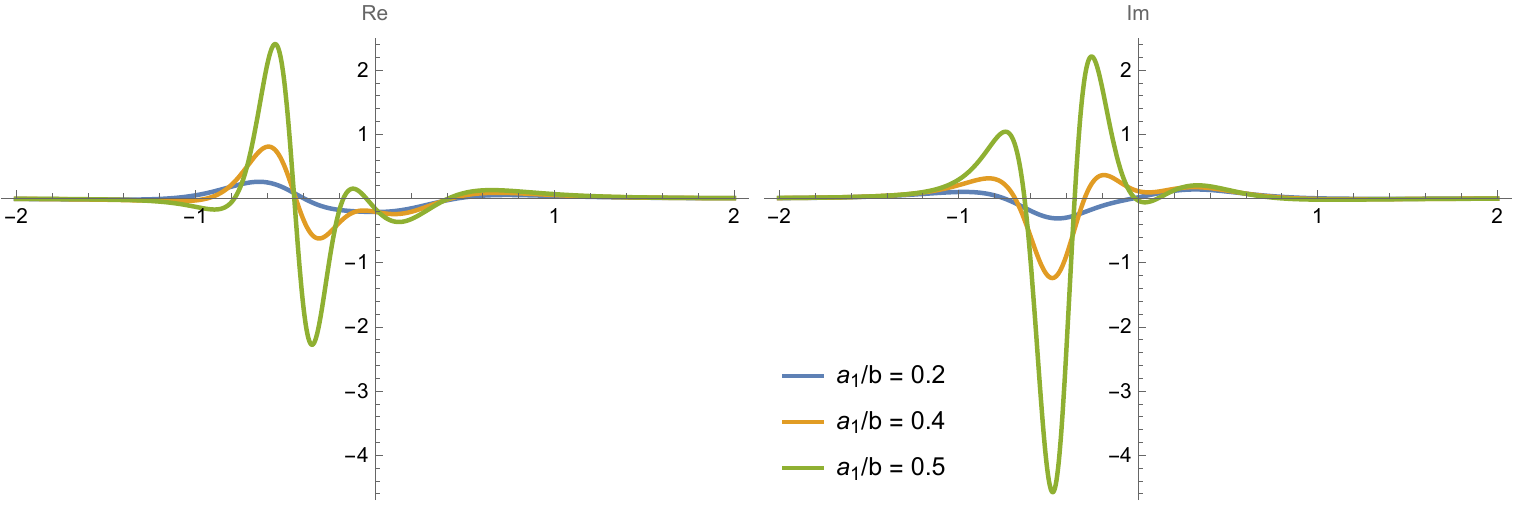}}
\end{center}
\caption{The $G_3$ waveform $(h^{\infty}_{+} + i  h^{\infty}_{\times})(u)$ plotted for $a_1\neq 0$, $a_2=0$. }
\label{fig:one-spinning-G3}
\end{figure}

\begin{figure}[ht]
\begin{center}
\scalebox{0.6}{\includegraphics{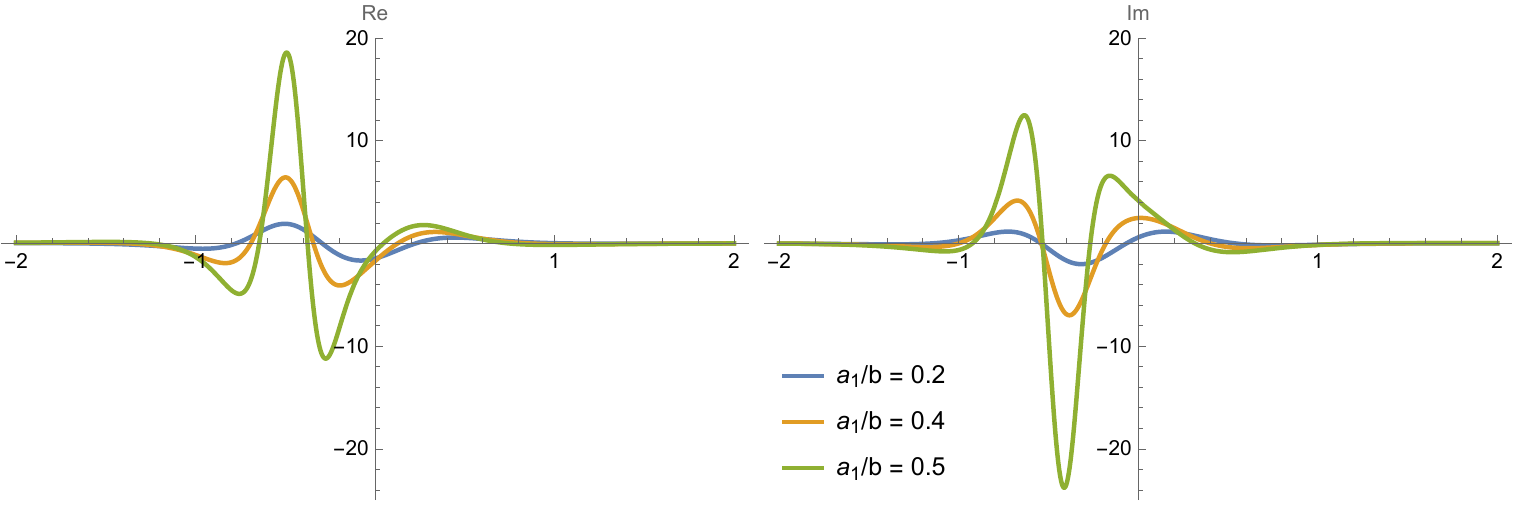}}
\end{center}
\caption{The $I_1$ waveform $(h^{\infty}_{+} + i  h^{\infty}_{\times})(u)$ plotted  for $a_1\neq 0$, $a_2=0$.}
\label{fig:one-spinning-I1}
\end{figure}

\begin{figure}[ht]
\begin{center}
\scalebox{0.6}{\includegraphics{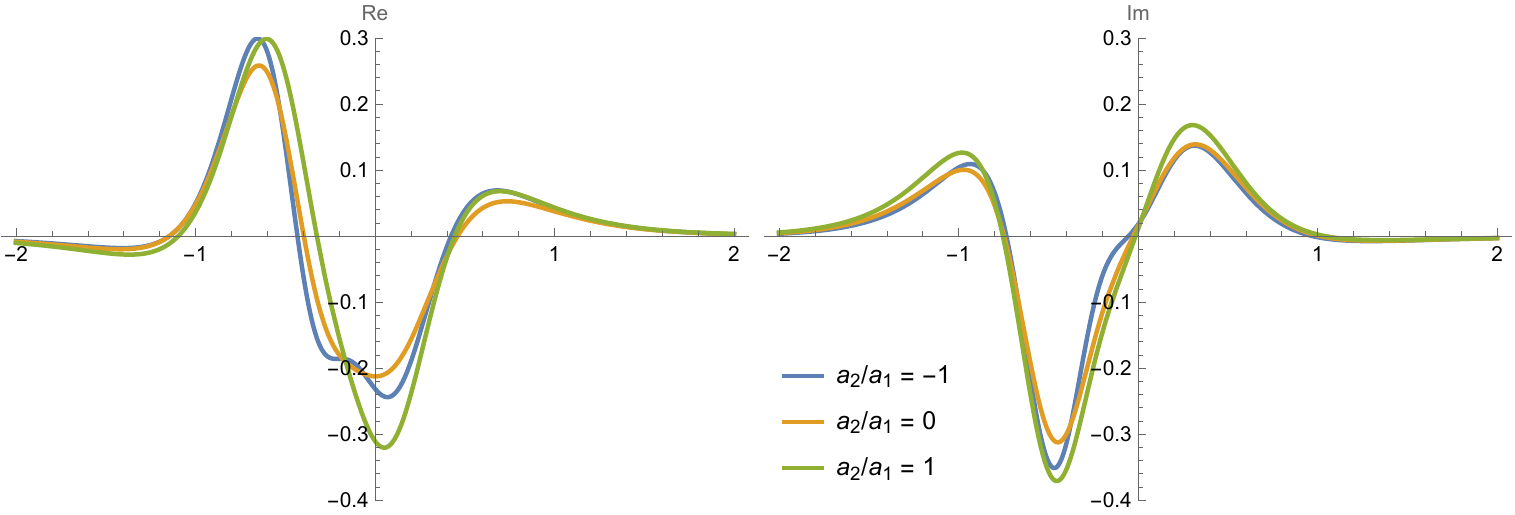}}
\end{center}
\caption{The $G_3$ waveform $(h^{\infty}_{+} + i  h^{\infty}_{\times})(u)$ plotted in the aligned spin case with $a_1/b {=} 0.2$ and  for various ratios $a_2/a_1$.}
\label{fig:spinning-G3}
\end{figure}

\begin{figure}[t]
\begin{center}
\scalebox{0.6}{\includegraphics{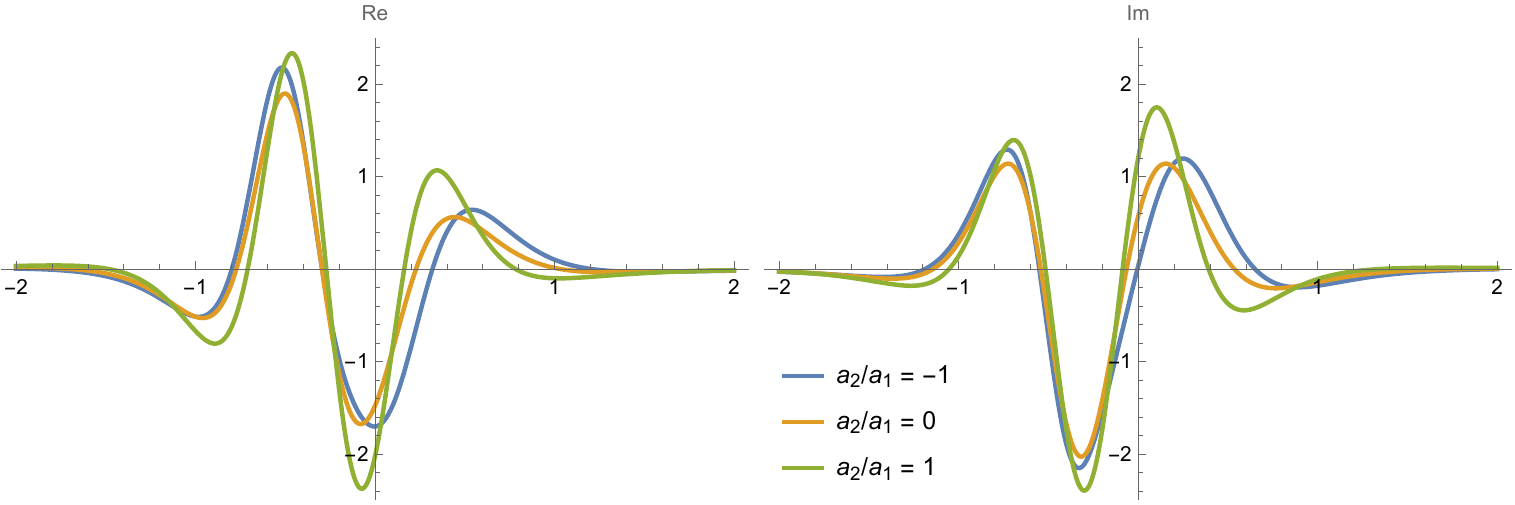}}
\end{center}
\caption{The $I_1$ waveform $(h^{\infty}_{+} + i  h^{\infty}_{\times})(u)$ plotted in the aligned spin case with $a_1/b{=}0.2$ and for various ratios $a_2/a_1$. Note that the amplitude of the $I_1$ deformations is about ten times larger than for the $G_3$ case.}
\label{fig:spinning-I1}
\end{figure}

We now show a few  $G_3$ and $I_1$ waveforms for  spinning objects. In Figures~\ref{fig:one-spinning-G3}~and~\ref{fig:one-spinning-I1} we consider the cases where only the spin of one of the two bodies is nonvanishing, say $a_1{\neq} 0$ and $a_2{=}0$, and  do so for increasing values of $a_1$.
In Figures~\ref{fig:spinning-G3} and~\ref{fig:spinning-I1} we show the $G_3$ and $I_1$  waveforms when both objects are spinning,  with their spins aligned, for various choices of the ratio $a_2/a_1$.
Note that the  gravitational memory is absent  also in the spinning case. 
We also recall that in setting $b{=}1$ we are measuring the spins $a_i$ in units of $b$. Thus, the values we are plotting here, e.g.~$a_i{=}\pm~0.5$, are rather large. Finally,  the waveforms also have poles at $b=\pm a_1\pm a_2$,% 
\footnote{These poles arise  from treating  the  spin dependence   exactly and are  also present  in the scattering  angle \cite{Brandhuber:2024bnz}.} although  these points are not problematic since we assume $|a_i|\leq  G m_i< b$.

%\FloatBarrier

\FloatBarrier

\section{(No) gravitational memory}
\label{sec:memory}

As it is well known, gravitational memory is related to soft limits of the five-point amplitude \cite{Strominger:2014pwa}, and a discussion of the memory in the General Relativity waveforms to all orders in spin was presented in \cite{Brandhuber:2023hhl,Aoude:2023dui}. We now wish to study if   cubic deformations alter in any way the gravitational memory. 
Defining the memory  as 
\begin{align}
    \Delta (h^{\infty}_{+} \pm i  h^{\infty}_{\times}) \coloneq 
    \left. (h^{\infty}_{+} \pm i  h^{\infty}_{\times})\right|_{u\to + \infty} - \left. (h^{\infty}_{+} \pm i  h^{\infty}_{\times})\right|_{u\to -  \infty}
    \, , 
\end{align}
one finds that (see e.g.~\cite{Herderschee:2023fxh, Brandhuber:2024bnz})
\begin{align}
\begin{split}
\label{mem-final}
    \Delta (h^{\infty}_{+} \pm i  h^{\infty}_{\times}) &= 
  -i \, \kappa S_{\rm W}^{\rm HEFT} \Big(\hat{k}, -i \frac{\partial}{\partial b}; \pm\Big)\, \delta_{\rm HEFT}
   \, ,  
\end{split}
\end{align}
where
\begin{align}
\label{deltaHEFT}
  \delta_{\rm HEFT}\coloneqq  
  \int\!d\mu^{(D)} e^{iq_\Cdot b} \, \big( -i \cM_4^{\rm HEFT}\big)(q)\, , 
\end{align}
is the Fourier transform to impact parameter space of the four-point classical amplitude $\cM_4^{\rm HEFT}$ \cite{Brandhuber:2021kpo,Brandhuber:2021bsf,Brandhuber:2021eyq}, which now has to be evaluated in the presence of cubic deformations; while 
and $S_{\rm W}^{\rm HEFT}$ is the classical (or HEFT) Weinberg soft factor   \cite{Brandhuber:2023hhy} 
\begin{align}
\begin{split}
	\label{SWHEFT}
	S_{\rm W}^{\rm HEFT} &= 
   - \frac{\kappa}{2}\,\frac{1}{\omega}\varepsilon_{\mu \nu}^{(h)}(k) \left[
		\frac{p_1^\mu q^\nu +p_1^\nu q^\mu}{p_1\Cdot \hat{k}} - p_1^\mu p_1^\nu \frac{q\Cdot \hat{k}}{(p_1\Cdot \hat{k})^2}\, - \, 1\leftrightarrow 2\right]
  \,  . 
\end{split}
\end{align}
 However, at leading order (tree level) in the presence of cubic deformations the classical four-point amplitude vanishes,  $\cM_4^{\rm HEFT, (0)}\!=\!0$; indeed,  this quantity  starts receiving corrections at one loop, or 2PM, which were computed  in \cite{Brandhuber:2024bnz}. We conclude that the leading-order gravitational memory in the presence of cubic deformations vanishes. This is also confirmed by the explicit plots shown in 
 the 
 %Figure~\ref{fig:no-memory}.
 figures  in Section~\ref{sec:pictures}. 

\section{Discussion}
\label{sec:conclusions}

In this paper we have considered  parity-even and -odd cubic deformations to general relativity, and computed analytically the corresponding corrections to the waveforms to leading order in the deformations and in Newton's constant. We have done so following three independent approaches, yielding  results in complete agreement. The direct integration method provides the most compact expressions, though the IBP reduction and residue methods are possibly more easily extendable beyond tree level. The residue method is especially   rewarding in that it easily shows  the absence of contributions from spurious poles. 

To conclude, we  briefly muse on the size  of the corrections due to the cubic deformations studied here and in \cite{Brandhuber:2024bnz}. Specifically, we   focus  on an observable quantity such as the impulse (or momentum kick) $\left.\Delta \mathbb{P}\right|_{R^3}$, and   ask when the corrections arising from cubic deformations become comparable to those  in general relativity at a certain PM order. 

Our cubic vertices have the form (schematically) $\int\!d^4x\, \beta \kappa^3 (\partial^2h)^3$. We  redefine the coupling constant in such a way that the three-point vertex scales as $\kappa$, thus introducing $\beta = \hat{\beta} / \kappa^2$; in terms of $\hat{\beta}$ the vertex has the form $\kappa \hat{\beta} \int\!d^4x\, (\partial^2h)^3$, and we  note that the dimension of $\hat{\beta}$ is $({\rm length})^4$, so that  we will set $\hat{\beta} = \ell_{\rm EFT}^4$. 
We found in \cite{Brandhuber:2024bnz} that, 
to leading order in the deformation and in $G$, the corrections to the impulse arising from cubic deformations have the form (see~(6.61)--(6.64) of that paper)
\begin{align}
\left.\Delta \mathbb{P}\right|_{R^3} \sim \beta \kappa^6 \frac{m^3}{b^6}\sim \frac{\hat\beta}{\kappa^2 }\frac{G^3 m^3}{b^6}\sim m \times \left(\frac{\ell_{\rm EFT}}{b}\right)^4 \left(\frac{Gm}{b}\right)^2 \, . 
\end{align}
On the other hand, to $n$PM order in general relativity the impulse  receives corrections    of the type 
\begin{align}
\left.\Delta \mathbb{P}\right|_{\rm EH, {\it n}}\sim m \times \left(\frac{Gm}{b}\right)^n
    \, , \end{align}
so that 
\begin{align}
\frac{\left.\Delta \mathbb{P}\right|_{R^3}}{\left.\Delta \mathbb{P}\right|_{\rm EH, {\it n}}} \sim \frac{\left(\dfrac{\ell_{\rm EFT}}{b}\right)^4}{\left(\dfrac{Gm}{b}\right)^{n-2}}\, . 
\end{align}
For hyperbolic encounters of interest we can choose for illustration  $G m/b \sim  1/5$, which ensures that  perturbation theory is still valid while also giving rise to a sizeable effect. By further requiring that, for observability,  ${\left.\Delta \mathbb{P}\right|_{R^3}}/ {\left.\Delta \mathbb{P}\right|_{\rm EH, {\it n}}} \sim 1$, we arrive at  the estimate 
\begin{align}
\label{boundd}
    \ell_{\rm EFT}\sim b \, (1/5)^{\frac{n-2}{4}}\, . 
\end{align}
As an   example, we can ask when   the corrections  from cubic deformations  become comparable to those arising at 5PM (or four loops)  in general relativity.%
\footnote{Parts of  the 5PM corrections to the impulse and the scattering angle for non-spinning objects were recently computed in \cite{Driesse:2024xad}.}
From  \eqref{boundd}, and further choosing $b\sim 100 \, {\rm km}$ we arrive at the estimate  $\ell_{\rm EFT} \sim 30 \, {\rm km}$, which is within the regime of validity of the effective field theory,
and also happens to be in agreement with the estimate   provided 
in~\cite{Silva:2022srr}. We also note that in \cite{Sennett:2019bpc}, bounds on the fundamental length scale  $\ell_{R^4}$ in a quartic effective field theory considered earlier in \cite{Endlich:2017tqa} were established, resulting in $\ell_{R^4} \lesssim 150\, {\rm km}$.

Similarly, we find that the waveform 
$h^\infty$ in general relativity at order $n$ in the PM expansion scales as
\begin{align}
h^\infty|_{\rm EH, {\it n}} \sim b \times \left( \frac{G m}{b}\right)^n \ ,
\end{align}
where now  $n{=}2$ corresponds to tree level and $n{=}3$  to one loop. 
For the leading cubic correction computed in this paper we found (see for example \eqref{G3-1} and \eqref{littlevomit})
\begin{align}
h^\infty|_{R^3} \sim b \times \left( \frac{G m}{b}\right)^2 \left( \frac{\ell_{\rm EFT}}{b}\right)^4 \ ,
\end{align}
leading to similar conclusions as above for the impulse \eqref{boundd}. Furthermore, if we assume that $\ell_{\rm EFT} \sim G m$ we can say that the effects of cubic interaction are in practice four orders higher in the PM expansion. This may be challenging given the currently available state of the art results in the PM expansion, but results in the PN framework might already give access to higher orders in $G$.

A final comment on the spin radii $a_i$. In the expressions for the impulse and waveform,  they appear through the shifted impact parameters (schematically) $\tilde{b} = b + a$ (omitting Lorentz indices and numerical factors). Recalling that for a physical Kerr black hole
$a/(G m) \leq 1$,  we can express this as
\begin{align}
\tilde{b} = b \left( 1 + \frac{a}{G m} \frac{G m}{b} \right) \ . 
\end{align}
This implies that effectively every factor of $a$ increases the  PM order by one,  since $GM/b$ is the effective loop counting parameter. For example,  the linear in spin part of our results  should be accompanied  by the one-loop correction of the scalar waveform in the cubic theory.
We hope to return to these questions in the  near future.

%\newpage
\section*{Acknowledgements}
We would like to thank  
Joshua Gowdy and Paolo Pichini 
for stimulating conversations.
GT thanks the Physics Department at the University of Rome ``Tor Vergata'' for their warm hospitality.   
This work was supported by the Science and Technology Facilities Council (STFC) Consolidated Grants   ST/T000686/1 and  ST/X00063X/1 \textit{``Amplitudes, Strings  \& Duality''}.
The work of GRB and PVM is supported by STFC quota studentships.   GC has received funding from the European Union's Horizon 2020 research and innovation program under the Marie Sk\l{}odowska-Curie grant agreement No.~847523 ``INTERACTIONS''. 
GT is also supported by a Leverhulme research fellowship RF-2023-279$\backslash 9$.
No new data were generated or analysed during this study.

\newpage

\appendix

\section{All tensor integrals we need}
\label{sec:AppA}

We summarise here relevant integrals needed to compute the time-domain waveforms in our paper. We focus here on integrals involving the $1/q_1^2$ propagator, while integrals involving the $1/q_2^2$ propagator can be  obtained by appropriate relabelling of the results quoted below. In the following we define $q^\mu \coloneq q_1^\mu$ and introduce the projector
\begin{align}
    P_1^{\mu\nu} \coloneq \eta^{\mu\nu} - v_1^\mu v_1^\nu \ .
\end{align}
Henceforth, we denote projected four-vectors  and their modulus as
\begin{align}
V^\mu_{(1)} \coloneq P_1^{\mu\nu} V_\nu \, , 
    \end{align}
    and 
    \begin{align}
        |V_{(1)}| \coloneq \sqrt{-V_{(1)} \Cdot V_{(1)} } = \sqrt{-V \Cdot P_1 \Cdot V}.
\end{align}
The first and simplest integral we encounter is
\begin{align}
\begin{split}
\label{motherofmanyintegrals}
\mathcal{I} & = \int \frac{d^4q}{(2 \pi)^3} \delta(2 q \Cdot v_1) \frac{e^{i q \cdot \tilde{b}}}{q^2}        \\
& = \int \frac{d^4q}{(2 \pi)^3} \delta(2 q \Cdot v_1) \frac{e^{i q \cdot \tilde{b}_{(1)}}}{q^2} \\
& = -\frac{1}{8 \pi |\tilde{b}_{(1)}|} \ ,
\end{split}
\end{align}
where in the second line we exploited the fact that due to the $\delta$-function we can insert for free the projector $P_1$ between $q$ and $\tilde{b}$ in the exponent. We actually need tensor versions of this integral, 
\begin{align}
\label{intoneden}
    \mathcal{I}^{\mu_1 \mu_2 \ldots \mu_n} = 
    \mathcal{I}[q^{\mu_1} q^{\mu_2} \cdots q^{\mu_n}] = 
    \int \frac{d^4q}{(2 \pi)^3} \delta(2 q\Cdot v_1 ) \, e^{i q \Cdot \tilde{b}}\ \frac{q^{\mu_1} q^{\mu_2} \cdots q^{\mu_n} }{q^2} \ ,
\end{align}
which can be simply obtained by acting  $n$ times on $\mathcal{I}$  with the derivative operator
\begin{align}
    -i \frac{\partial}{\partial \tilde{b}_{(1) \mu_i}} \ ,
\end{align}
and using the identity 
\cite{Aoude:2023dui} 
\begin{align}
    \frac{\partial \tilde{b}_{(1)}^\mu}{\partial \tilde{b}_{(1)\nu}} = P_1^{\mu\nu} \ .
\end{align}
Two examples are 
\begin{align}\label{Imu}
    \cI^\mu = -i \frac{\partial \mathcal{I}}{\partial \tilde{b}_{(1)\mu}} = \frac{i}{8 \pi} \frac{\tilde{b}_{(1)}^\mu}{|\tilde{b}_{(1)}|^3} \, , \end{align}
    and 
    \begin{align}
    \label{Imunusimple}
        \cI^{\mu\nu} = - \frac{\partial^2 \mathcal{I}}{\partial \tilde{b}_{(1)\mu} \partial \tilde{b}_{(1)\nu}} = \frac{|\tilde{b}_{(1)}|^2 P_1^{\mu\nu}+3 \tilde{b}^\mu_{(1)} \tilde{b}^\nu_{(1)}}{8 \pi |\tilde{b}_{(1)}|^5}\ .
\end{align}
In a similar fashion, we can obtain the other two more complicated tensor integral families. 
The first one is 
\begin{align}
    \mathcal{J^\mu} \coloneq \int \frac{d^4q}{(2 \pi)^3} \delta(2 q \Cdot v_1) e^{i q \cdot \tilde{b}}\frac{ q^\mu}{q^2 \ \hat{k} \Cdot q} \ ,
\end{align}
which, as already mentioned at the end of Section~\ref{ssec:I1-wf}, requires a regulator because of the $1/\hat{k} \Cdot q$ pole.
This is a spurious pole that cancels in the sum of the two diagrams contributing to  the waveform and, of course, the final result is independent of the regulator. For  illustration we will present here  the evaluation of $\mathcal{J}^\mu$ using both the principal value (PV) and $i\varepsilon$ prescriptions, while in the main text we  only make use the PV prescription.

We begin discussing  the PV prescription.
We  denote the PV regulated integral by $\mathcal{J}^\mu$,  which can  be expressed in terms of the scalar integral 
\begin{align}
\label{Jmu-def}
    \mathcal{J} \coloneq \int \frac{d^4q}{(2 \pi)^3} \delta(2 q \Cdot v_1) \frac{e^{i q \cdot \tilde{b}}}{q^2 \ \hat{k} \Cdot q} \ ,
\end{align}
as
\begin{align}
    \mathcal{J}^\mu = -i \frac{\partial \mathcal{J}}{\partial \tilde{b}_{(1)\mu}} \ .
\end{align}
Following \cite{Jakobsen:2021lvp},  we now notice that  $\mathcal{J}$ is invariant under rescaling of $\tilde{b}$. Indeed a rescaling $\tilde{b} \to \alpha \, \tilde{b}$ can be undone by the rescaling $q \to q/\alpha$ which leaves the integrand unchanged. Hence, we infer that 
\begin{align}
    \tilde{b}_{(1)\mu} \mathcal{J}^\mu = -i \, \tilde{b}_{(1)\mu} \frac{\partial \mathcal{J}}{\partial \tilde{b}_{(1)\mu}} = 0 \ , 
\end{align}
and we further notice that $v_{1 \mu} \mathcal{J}^\mu = 0$ since $v_1 \Cdot q = 0$ due to the $\delta$-function in the integral. Hence $\mathcal{J}^\mu$ must live in a two-dimensional subspace orthogonal to $v_1^\mu$ and $\tilde{b}_{(1)}^\mu$. In order to implement this,  it is convenient  to introduce the symmetric tensor~\cite{Jakobsen:2021lvp}
\begin{align}
\label{K1def}
K_1^{\mu\nu} = |\tilde{b}_{(1)}|^2 P_1^{\mu\nu} + \tilde{b}_{(1)}^\mu 
\tilde{b}_{(1)}^\nu \ ,
\end{align}
obeying $K_1^{\mu\nu} v_{1\nu} = K_1^{\mu\nu} \tilde{b}_{(1)\nu} = 0$. This allows us to write an ansatz for $\mathcal{J}^\mu$ as
\begin{align}
\mathcal{J}^\mu = c \, K_1^{\mu\nu} \hat{k}_\nu \ .    
\end{align}
We can now solve for $c$ by observing that 
\begin{align}
    \mathcal{J}^\mu \hat{k}_\mu = \mathcal{I} = -\frac{1}{8 \pi |\tilde{b}_{(1)}|} = c \, \hat{k} \Cdot K_1 \Cdot \hat{k} \, , 
    \end{align}
    from which it follows that 
    \begin{align} c = -\frac{1}{8 \pi |\tilde{b}_{(1)}| \hat{k} \Cdot K_1 \Cdot \hat{k}} \ .
\end{align}
Therefore, we find that 
\begin{align}
\label{Jmu}
\mathcal{J}^\mu = \frac{K_1^{\mu\nu} \hat{k}_\nu}{8 \pi |\tilde{b}_{(1)}| \, \big[(v_1 \Cdot \hat{k})^2 |\tilde{b}_{(1)}|^2-(\hat{k} \Cdot \tilde{b}_{(1)})^2 \big]}
 \ ,
\end{align}
and from this result we can obtain $\mathcal{J}^{\mu \mu_1 \ldots \mu_n}$ with any number of extra $q^{\mu_i}$ insertions by taking derivatives with respect to $\tilde{b}_{(1) \mu_i}$ as explained above.

In order to obtain tensor integrals,  the form above is preferred as it makes the $\tilde{b}_{(1)}$-dependence fully transparent. On the other hand, one can obtain a more compact expression  in terms of the shifted, unprojected impact parameter $\tilde{b}$ as follows. 
First, it is important to note that $\tilde{b} \Cdot \hat{k}=0$ as assumed throughout the paper. Note that any shift of $\tilde{b}$ by $\tilde{a}_i$
preserves this property
since $\tilde{a}_i \Cdot \hat{k}=0$.
Under this assumption we can then write 
\begin{align}
\tilde{b}^\mu = \tilde{b}_{(1)}^\mu - \frac{\hat{k} \Cdot \tilde{b}_{(1)}}{\hat{k} \Cdot v_1} v_1^\mu \, , 
\end{align}
from which it follows that 
\begin{align}
  |\tilde{b}\, |^2 \coloneq-\tilde{b} \Cdot \tilde{b} =  |\tilde{b}_{(1)}|^2 - 
\frac{(\hat{k} \Cdot \tilde{b}_{(1)})^2}{(\hat{k} \Cdot v_1)^2} \ . 
\end{align}
We  can then  rewrite  $\mathcal{J}^\mu$ in the more compact form (if $\tilde{b} \Cdot \hat{k}=0$)
\begin{align}
    \mathcal{J}^\mu =  \frac{K_1^{\mu\nu} \hat{k}_\nu}{8 \pi  (v_1 \Cdot \hat{k})^2 |\tilde{b}_{(1)}| \, |\tilde{b}\, |^2}
 \ .
\end{align}
Next, we examine  the same integral,  this time  regulated with an $i\varepsilon$ prescription:
\begin{align}
\begin{split}
\label{Jmu-def-eps}
    \mathcal{J}_{\varepsilon}^\mu & \coloneq \int \frac{d^4q}{(2 \pi)^3} \delta(2 q \Cdot v_1) \frac{q^\mu e^{i q \cdot \tilde{b}}}{q^2 \ (\hat{k} \Cdot q - i \varepsilon)} \\
 & =   i \int_{-\infty}^0 d\tau
 \int \frac{d^4q}{(2 \pi)^3} \delta(2 q \Cdot v_1) \frac{q^\mu e^{i q \cdot (\tilde{b} + \tau \hat{k})  + \tau \varepsilon}}{q^2}
    \end{split} \\
    & = -\frac{1}{8 \pi}\int_{-\infty}^0 d\tau 
    \frac{\tilde{b}^\mu_{(1)} + \tau \hat{k}^\mu_{(1)}}{\left[ -(\tilde{b}+\tau \hat{k}) \Cdot P_1 \Cdot (\tilde{b}+\tau \hat{k})\right]^{3/2}} \ ,
\end{align}
where in the second line we have used the Schwinger trick and in the last line we have used \eqref{Imu}. Performing the $\tau$ integration gives
\begin{align}
\mathcal{J}_{\varepsilon}^\mu =
\frac{|\tilde{b}_{(1)}| \hat{k}^\mu_{(1)} - v_1 \Cdot \hat{k} \, \tilde{b}^\mu_{(1)}}{8 \pi (v_1 \Cdot \hat{k}) |\tilde{b}_{(1)}| (|\tilde{b}_{(1)}| v_1 \Cdot \hat{k} + \tilde{b}_{(1)} \Cdot \hat{k} )} \ ,
\end{align}
and similarly, replacing $i \varepsilon \rightarrow -i \varepsilon$ we find
\begin{align}
\mathcal{J}_{-\varepsilon}^\mu =
\frac{|\tilde{b}_{(1)}| \hat{k}^\mu_{(1)} + v_1 \Cdot \hat{k} \, \tilde{b}^\mu_{(1)}}{8 \pi (v_1 \Cdot \hat{k}) |\tilde{b}_{(1)}| (|\tilde{b}_{(1)}| v_1 \Cdot \hat{k} - \tilde{b}_{(1)} \Cdot \hat{k} )} \ .
\end{align}
It can then be checked that these integrals are related to the PV integral \eqref{Jmu} as
\begin{align}
\mathcal{J}^\mu = \frac{1}{2} \left( \mathcal{J}^\mu_\varepsilon + \mathcal{J}^\mu_{-\varepsilon}\right) \ .
\end{align}
Finally, we tackle (with PV regularisation) the most complicated integral,
\begin{align}
\label{Kmunu-defint}
\mathcal{K}^{\mu\nu} = \int\! \frac{d^4q}{(2 \pi)^3} \delta(2 v_1 \Cdot q)
\frac{e^{i q \Cdot \tilde{b}} q^\mu q^\nu}{q^2 \ \hat{k} \Cdot q \ a_2 \Cdot q_2} 
= \int\!\frac{d^4q}{(2 \pi)^3} \delta(2 v_1 \Cdot q)
\frac{e^{i q \Cdot \tilde{b}} q^\mu q^\nu}{q^2 \ \hat{k} \Cdot q \ (- \tilde{a}_2 \Cdot q)}\ ,
\end{align}
where 
we have introduced the shifted spin radii 
\begin{align}
    \tilde{a}_i^\mu = a_i^\mu - \frac{a_i \Cdot \hat{k}}{v_i \Cdot \hat{k}} v_i^\mu \, ,  \qquad \tilde{a}_i \Cdot \hat{k} = 0 \, ,
\end{align}
for $i=1,2$.
Also note that $q_2^\mu = k^\mu - q_1^\mu = \omega^* \hat{k}^\mu - q^\mu = \frac{v_2 \Cdot q}{v_2 \Cdot \hat{k}} \hat{k}^\mu - q^\mu$, and therefore we can rewrite $a_2 \Cdot q_2 = - \tilde{a}_2 \Cdot q$.
$\mathcal{K}^{\mu\nu}$ has the following important properties:
\begin{align}
    \mathcal{K}^{\mu\nu} v_{1\nu} = \mathcal{K}^{\mu\nu} \tilde{b}_{\nu}
    = \mathcal{K}^{\mu\nu} \tilde{b}_{(1)\nu} =\mathcal{K}^{\mu\nu}  \eta_{\mu\nu} = 0\, .
\end{align}
The first is due to the $\delta$-function, the second and third property are consequence of the rescaling invariance of $\mathcal{K}^\mu$, and the fourth property (tracelessness) can be seen as follows: taking the trace and
choosing to work in the rest frame $v_1=(1,0,0,0)$, we find
\begin{align}\label{spurious-integrals}
\begin{split}
\mathcal{K}^{\mu\nu}  \eta_{\mu\nu}  & = \int\!\frac{d^4q}{(2 \pi)^3} \, \delta(2 q\Cdot v_1 )
\frac{e^{i q \Cdot \tilde{b}}}{\hat{k} \Cdot q \, (- \tilde{a}_2 \Cdot q)} \\
& = -\int_{-\infty}^0\!dt_1 dt_2 \int\!\frac{d^3\vec{q}}{2 (2 \pi)^3} \, e^{-i \vec{q} \Cdot (\vec{\tilde{b}}+t_1 \vec{\hat{k}} + t_2 \vec{\tilde{a}}_2)} \\
& = -\frac{1}{2} \int_{-\infty}^0 dt_1 dt_2 \ \delta^{(3)}(\vec{\tilde{b}}+t_1 \vec{\hat{k}} + t_2 \vec{\tilde{a}}_2) \ ,
\end{split}
\end{align}
which vanishes for generic choices of the vectors.

Due to the properties mentioned above, the integral \eqref{Kmunu-defint} must be a symmetric tensor living in the two-dimensional subspace orthogonal to $v_1$ and $\tilde{b}$, 
\begin{align}
    \mathcal{K}^{\mu\nu} = c_1 K_{1}^{\mu\nu} + 2 c_2 (\tilde{a}_2 \Cdot K_{1})^{(\mu} (\hat{k} \Cdot K_{1})^{\nu)} \ ,
\end{align}
where \begin{align}
    K_{1}^{\mu\nu} \coloneq |\tilde{b}_{(1)}|^2 P_1^{\mu\nu} + \tilde{b}_{(1)}^\mu \tilde{b}_{(1)}^\nu\, . 
    \end{align}
    Notice that no other structures are allowed in this ansatz, since the result must be symmetric under swapping $\hat{k} \leftrightarrow -\tilde{a}_2$ and be rescaled by $1/(\alpha_1 \alpha_2)$ if we perform the replacement  $\{\hat{k}, \tilde{a}_2\} \rightarrow \{\alpha_1 \hat{k}, \alpha_2 \tilde{a}_2\}$. 
Requiring also tracelessness, we instantaneously find
$c_1 = -c_2 \big( \tilde{a}_2 \Cdot K_{1} \Cdot \hat{k} \big)$.
On the other hand
\begin{align}
    \mathcal{K}^{\mu\nu} \hat{k}_\mu \tilde{a}_{2\nu} = - \mathcal{I} = \frac{1}{8 \pi |\tilde{b}_{(1)}|} = c_2 \left( \tilde{a}_2 \Cdot K_{1} \Cdot \tilde{a}_2 \right)\big( \hat{k} \Cdot K_{1} \Cdot \hat{k} \big) = - c_2 \left( \tilde{a}_2 \Cdot K_{1} \Cdot \tilde{a}_2 \right) (v_1 \Cdot \hat{k})^2 |\tilde{b}|^2 \ , 
\end{align}
and hence we find
\begin{align}
\label{Kmunu}
    \mathcal{K}^{\mu\nu} = -\frac{\big( \tilde{a}_2 \Cdot K_{1} \Cdot \hat{k} \big) K_{1}^{\mu\nu} -2 (\tilde{a}_2 \Cdot K_{1})^{(\mu} (\hat{k} \Cdot K_{1})^{\nu)}}{8 \pi  |\tilde{b}_{(1)}| \left( \tilde{a}_2 \Cdot K_{1} \Cdot \tilde{a}_2 \right) \big( \hat{k} \Cdot K_{1} \Cdot \hat{k} \big)} \ .
\end{align}
If we further assume $\tilde{b} \Cdot \hat{k}=0$, which is
true for all possible shifted versions of the impact parameter $\tilde{b} \pm i \tilde{a}_1 \pm i \tilde{a}_2$,
we find
\begin{align}
\label{Kmunu-simp}
    \mathcal{K}^{\mu\nu} = \frac{\big( \tilde{a}_2 \Cdot K_{1} \Cdot \hat{k} \big) K_{1}^{\mu\nu} -2 (\tilde{a}_2 \Cdot K_{1})^{(\mu} (\hat{k} \Cdot K_{1})^{\nu)}}{8 \pi (v_1 \Cdot \hat{k})^2 |\tilde{b}|^2 |\tilde{b}_{(1)}| \left( \tilde{a}_2 \Cdot K_{1} \Cdot \tilde{a}_2 \right) } \ .
\end{align}
Finally, we comment on potential terms in the waveform integrand that do not have physical poles $1/q_i^2$ with $i=1,2$, but instead have a pole of the form $1/(w \Cdot q)$ (or a product of such factors). In such cases the same manipulation as in \eqref{spurious-integrals} guarantees that such terms make vanishing contributions to the waveform. In  other words,  the tracelessness of integrals like $\mathcal{K}^{\mu \nu}$ in \eqref{Kmunu-defint} guarantees that terms with only spurious poles do not contribute to the waveform. This fact was also noticed in the Einstein-Hilbert case in \cite{Aoude:2023dui}. For this reason, it is not necessary to use the full five-point amplitude,  but the sum of the two factorisation channels \eqref{2factordiagrams} suffices. Similarly, all polynomial terms that only give rise to (derivatives of) $\delta$-functions can be ignored.

\section{Tensor integrals from generating functions}
\label{sec:AppB}
We now describe an alternative (and faster) way to compute tensor integrals. All of the integrals we are interested in  have the form
\begin{align}\label{integraldotprod}
\begin{split}
\cI_{m_1\ldots m_n}^{ [\cD]} (\tilde{b})  & \coloneq \int\!\frac{d^4q}{(2 \pi)^3} \, \delta(2 q\Cdot v_1 )\, 
e^{i q \Cdot \tilde{b}}\, \frac{(q_1\Cdot c_1)^{m_1} \cdots  (q_1\Cdot c_n)^{m_n}}{\cD} 
 \ ,
\end{split}
\end{align}
for some denominator $\cD$, and where $c_i$, $i=1, \ldots , n$ are four-vectors. These can be obtained very easily from the integral 
\begin{align}\label{integralt}
\begin{split}
\cI^{[\cD]} (\tilde{b}; t_1, \ldots, t_n) & \coloneq \int\!\frac{d^4q}{(2 \pi)^3} \, \delta(2 q\Cdot v_1 )\, 
 \frac{e^{i q \Cdot \big( \tilde{b} + \sum_{i=1}^n t_i c_i\big)}}{\cD} \\ 
 & = \cI^{[\cD]} (\tilde{b} + \sum_{i=1}^n t_i c_i; 0, \ldots, 0)
  \ ,
\end{split}
\end{align}
as
\begin{align}
    \cI_{m_1\ldots m_n}^{[\cD]}   = \left.(-i)^{\sum_{i=1}^n m_i} \frac{\partial^{\sum_{i=1}^n m_i}}{\partial^{m_1}t_1 \cdots \partial^{m_n}t_n}\cI^{[\cD]} (\tilde{b}; t_1, \ldots, t_n)\right|_{t_1 = t_2 = \cdots = t_n = 0}\, .
\end{align}
For instance, for the case of the $G_3$ waveform we need an integral of the form 
\begin{align}
\label{G3int}
\begin{split}
\int\!\frac{d^4q}{(2 \pi)^3} \, \delta(2 q\Cdot v_1 )\, 
e^{i q \Cdot \tilde{b}}\, \frac{(q \Cdot v_2)^{2}(q \Cdot X)^{2} }{q^2} \, , 
\end{split}
\end{align}
where $X^\mu\coloneq[\hat{k}| v_1 
\sigma^\mu | \hat{k}]$. This can be evaluated as 
\begin{align}
\label{G3intderiv}
    \begin{split}
&\left. \frac{\partial^{4}}{\partial^{2}t_1  \partial^{2}t_2}\int\!\frac{d^4q}{(2 \pi)^3} \, \delta(2 q\Cdot v_1 )\, 
 \frac{e^{i q \Cdot \big( \tilde{b} + t_1v_2 + t_2 X\big)}}{q^2} \right|_{t_1 = t_2  
= 0}
 \\ & 
= -\frac{1}{8 \pi}\left(\frac{\partial^{4}}{\partial^{2}t_1  \partial^{2}t_2}
\frac{1}{
|\tilde{b}_{(1)} + t_1 v_{2(1)} + t_2 X_{(1)}|}\right)_{t_1 = t_2 = 0} 
 \ ,
\end{split}
\end{align}
where we used \eqref{motherofmanyintegrals}, and as usual we adopt the notation defined in \eqref{V(1)} for projected vectors (but note that $X_{(1)} = X)$.

Summarising, in this approach we can easily relate tensor integrals to simpler ones (not necessarily scalar integrals, as seen in the previous section), without the need to perform Lorentz contractions and with the advantage of having to perform differentiations with respect to scalar parameters rather than vectors. 

\newpage

\bibliographystyle{JHEP}
\bibliography{ScatEq.bib}

\end{document}